\documentclass[fleqn,usenatbib]{mnras}

\usepackage[T1]{fontenc}
\usepackage{ae,aecompl}

\usepackage{graphicx}	
\usepackage{epsfig}
\usepackage{epsf}
\usepackage{amsmath}	
\usepackage{amssymb}	
\usepackage{xcolor}
\usepackage{ulem}
\usepackage{verbatim}
\usepackage[flushleft]{threeparttable}
\hypersetup{draft}

\newcommand{\CI}[0]{{}C\,{\sc i}\,}
\newcommand{\OI}[0]{{}O\,{\sc i}\,}
\newcommand{\CII}[0]{{}C\,{\sc ii}\,}
\newcommand{\SiII}[0]{{}Si\,{\sc ii}\,}

\newcommand{\UV}[0]{{}$I_{\rm{UV}}$}
\newcommand{\nH}[0]{{}${\rm{n_H}}$}

\title[H$_2$/\CI\ in local and high redshift galaxies]
{Physical conditions in diffuse interstellar medium of local and high redshift galaxies: measurements based on excitation of H$_2$ rotational and C\,{\sc i} fine-structure levels.}

\author[V.V.~Klimenko \& S.A.~Balashev]{V.V.~Klimenko\,$^{1}$ and S.A.~Balashev\,$^{1}$
	\\
	$^{1}$Ioffe Institute, {Polytekhnicheskaya ul. 26}, 194021 Saint Petersburg, Russia -- email: s.balashev@gmail.com
}

\date{\today}

\pubyear{2019}

\begin{document}
\label{firstpage}
\pagerange{\pageref{firstpage}--\pageref{lastpage}}
\maketitle

\begin{abstract}
	We present results of analysis of physical conditions (number density, intensity of UV field, kinetic temperature) in the cold H$_2$-bearing interstellar medium of local and high redshift galaxies. Our measurements based on the fit to the observed population of H$_2$ rotational levels and \CI\ fine-structure levels with the help of grids of numerical models calculated with the PDR Meudon code. A joint analysis of low H$_2$ rotational levels and \CI\ fine-structure levels allows to break the degeneracy in the $I_{UV}-n_{\rm{H}}$ plane and provides significantly tighter constraints on the number density and intensity of UV field. 
	Using archive data from the VLT/UVES, KECK/HIRES. HST/STIS and FUSE telescopes we selected 12 high redshift  damped Ly$\alpha$ systems (DLAs) in quasar spectra and 14 H$_2$ absorption systems along the lines of sight towards stars in the Milky-Way and the Magellanic Clouds galaxies. These systems have strong H$_2$ components with the column density $\log N({\rm{H_2}})/{\rm{cm}}^{-2}>18$ and associated \CI\ absorptions. We find that H$_2$-bearing medium in high redshift DLAs and in local galaxies has similar values of the kinetic temperatures $T_{\rm{kin}}\sim100$\,K and number density $10-500$\,cm$^{-3}$. However, the intensity of incident UV radiation in DLAs is varied in the wide range ($0.1-100$ units of Mathis field), while it is $\sim0.1-3$ units of Mathis field for H$_2$ systems in Milky-Way and LMC and SMC galaxies.
    The large dispersion of measured UV flux in DLAs is probably a consequence that DLA sample probes the galaxies selected from the overall galaxy population at high-z and therefore corresponds to the wide range of the physical conditions.
 
\end{abstract}




\section{Introduction}
\label{sec:intro}

The absorption line analysis towards background sources is a powerful method to study local physical conditions in the interstellar medium (ISM). It has some advantages over emission line studies, in particular, it provides column density measurements over very small transverse scales. In the local Universe this corresponds to the extra tiny radius of stars, but even for remote Universe the sizes of emitting region of quasars ($< 1$\,pc) and gamma-ray burst afterglows  are significantly smaller than typical spatial resolution of emission line studies. Additionally, high resolution spectra enable to resolve absorption lines and therefore determine column densities with very high accuracy ($<0.1$ dex). Another important aspect, that absorption line studies allow us to directly probe H$_2$ (along with many other associated species) through resonant rest-frame UV lines, the so-called Lyman and Werner bands. This leads that this technique is a perfectly suitable to study diffuse phases of the cold ISM, which are very hardly accessible through emission studies, even in our own galaxy. This include so-called "CO-dark" gas which can be the dominant form of the molecular gas at low metallicities \citep{Balashev2017}. Usually, the optical and especially UV wavelength range is suitable for the absorption line studies, since most of resonant permitted electronic transitions are located in these ranges. For the high redshift measurement it provides even more advantage, since UV range due to cosmological redshift is shifted in optical domain, which allows to obtain high resolution spectra at the largest ground-based optical telescopes \citep{Noterdaeme2008}. 

To determine {\sl local} physical conditions one can use the measurements of the relative population of the energy levels of the different species. For the cold diffuse ISM, with typical temperatures $\lesssim100$\,K, the most appropriate energy systems are rotational levels of H$_2$, HD \citep[e.g.][]{Balashev2010} and fine-structure levels of \CI, \CII, \SiII\ and \OI\ \citep[see e.g.][]{Silva2002}. 
However, \OI\ has a relatively large distance between the fine-structure levels, and therefore lines from excited levels of \OI\ is observed very seldom. Additionally, usually, it is very hard to constrain the column density even for the ground level of \OI, since available resonant UV lines are usually saturated. While the fine-structure excitation of \SiII\ is not very often observed in DLAs in contrast to \CII, they were both analysed in several DLAs \citep{Srianand2000a, Howk2005, Kulkarni2012, Neeleman2015}. However for both \SiII\ and \CII\ even if available always correspond to the mix of the cold and warm phases \citep[e.g.][]{Neeleman2015}. 
In contrast to \OI\ and \SiII, the excitation of \CI\ fine-structure levels is a perfect tracer of the cold diffuse ISM. Historically, it was used to measure the thermal pressure in the cold ISM in the Milky Way \citep[MW,][]{Jenkins2011}, Magellanic clouds \citep[MC,][]{Welty2016} and in low \citep{Muzahid2016} and high redshift DLAs \citep{Srianand2005, Jorgenson2010, Balashev2019}. 

The population of H$_2$ levels also is perfectly suitable for the determination of the physical condition of the cold diffuse ISM. Such measurements, usually requires the computational expensive modelling of the cold ISM including detailed radiative transfer in resonant H$_2$ lines, which are typically in optically thick regime. The latter results in that excitation of H$_2$ levels are highly homogeneous within the cloud \citep{Abgrall1992, Balashev2009}. However, for saturated H$_2$ absorption systems, with $\log N({\rm{H_2}})>18$ (where $N$ is column density measured in cm$^{-2}$), 
the levels of J=0,1,2 are predominantly thermalized and their excitation is typically close to the thermal temperature \citep{LePetit2006}, which is set by the thermal balance, itself being a function of the density and UV field. 
Indeed, the main heating mechanism in the diffuse neutral ISM is a photoelectric effect on grains, whereas cooling is a combination of the emission in C\,{\sc ii} and O\,{\sc i} lines and electron recombination \citep{Draine2011}. Rates of these processes are strongly dependent on the gas density, metallicity and intensity of UV radiation. 
On the other hand, the excitation of higher rotational levels of H$_2$ at J$\ge3$ is determined mainly by the UV pumping \citep{Black1976} and regulated by the intensity of UV radiation. Therefore, populations of both low and high rotational levels of H$_2$ carry information on the physical conditions, that makes them a promising tool for estimation physical parameters in the diffuse ISM.

However, with aforementioned difficulties, H$_2$ was previously used to estimate physical condition (number density and UV intensity) only in the specific cases in interstellar clouds in the Milky-Way \citep[e.g.][]{Jura1975,Jura1975b,Nehme2008} and in a few high-redshift H$_2$-bearing DLAs, whose analysis includes detailed and complex modeling of a single system  \citep{Noterdaeme2007, Klimenko2016, Shaw2016, Noterdaeme2017, Rawlins2018, Shaw2020}.

In this paper we present a systematic study of physical conditions in the cold ISM of high redshift and local galaxies based on the analysis of excitation of the rotational levels of H$_2$ and the fine-structure levels of \CI. For these we used all available detected high redshift DLAs and sample of absorption systems in stellar spectra of the Milky-Way and Magellanic Clouds. The structure of the paper is following. In Sect.~\ref{sec:method} we describe a method to analyse physical conditions of molecular gas using observed populations of H$_2$ rotational levels and fine-structure levels of \CI.  The data of known H$_2$ and \CI\ absorption systems in high redshift DLAs and local galaxies is compiled in Sect.~\ref{sec:sample}. In Sect.~\ref{sec:results} we present our measurements of the number density, kinetic temperature and intensity of the incident UV field in the samples. Then we discuss the dependencies of the inferred parameters and compare the measurements in local Universe and high redshift DLAs, before we conclude in Sect.~\ref{sec:conclusion}.

\section{Method}
\label{sec:method}

The excitation of most H$_2$ levels in the cold diffuse ISM is determined predominantly by the UV pumping that takes place in resonant UV lines. At typical H$_2$ column densities in observed absorption systems ($\log N (\rm H_2) \gtrsim 16$), the UV lines can be significantly saturated, and hence, the excitation strongly varied within the cloud. Therefore the modelling of the excitation of H$_2$ levels requires an accurate calculation of radiative transfer to take into account shielding of H$_2$ absorption lines. Additionally, the population of the lower H$_2$ rotational levels ($J\le2$) can depend on the temperature of the gas, which also can be varied within the cloud, due to changes in the chemical state (e.g. HI/H$_2$ transition) and dilution of the UV field. Therefore we used the {\sc PDR Meudon code}\footnote{We use a version 1.5.4 rev 2053 (from 03 March 2020)} \citep{LePetit2006}, which performs a complete calculation of the radiative transfer of UV radiation in the UV lines of H$_2$ in combination with a solution of the thermal balance and chemistry. 
A model consists of a slab cloud of the gas irradiated by isotropic interstellar radiation field from both sides. We adopted the standard model of the interstellar radiation (ISRF) proposed by \cite{Mathis1983}. This ISRF covers a large wavelength range, from 912\,\AA\, to sub-millimeter. The intensity of the ISRF in the UV range was scaled by a factor \UV. That is, $I_{\rm{UV}}=1$ corresponds to one unit of Mathis field or 0.75 unit of the Draine field \citep{Draine1978} in the interval between 912\,\AA\,-1100\,\AA\,. The cosmic microwave background (CMB) radiation is also the important component of the ISRF. The direct excitation by the CMB photons can be dominating mechanism of excitation of the first fine-structure level of \CI\ in diffuse gas at high redshift \citep[e.g.][]{Silva2002}. Therefore we used two set of PDR models calculated with different CMB temperatures $T_{\rm CMB}=2.73\times(1+z)$ at z=0 for comparison with data samples of MW and MC measurements ($T_{\rm CMB}=2.73$\,K), and at average redshift $z=2.5$ for comparison with a sample of high redshift systems ($T_{\rm CMB}=9.56$\,K). The full range of the redshift of high redshift DLAs sample is $2.0<z_{\rm abs}<3.3$. The difference of the redshifts between absorption systems and the models introduces a systematic bias of the estimate of the number density (or thermal pressure). 
However this bias rapidly vanish with an increase of the number density, since collisional excitation starts to dominate in \CI\ populations. We checked that for the studied systems this bias is not significantly larger than 0.2~dex, which is the typical statistical uncertainty of the present method. 

The size of the modelled cloud was defined by the visual extinction $A_{\rm V}^{\rm max}$ parameter. We fixed it to $A_{\rm V}^{\rm max}=0.5$ that corresponds to the total hydrogen column density $N({\rm H})_{\rm tot} \simeq 10^{21}/Z\mbox{\,cm}^{-2}$, where $Z$ is the metallicity. This maximal value exceeds the typical visual extinctions measured at high-z DLAs \citep[e.g.][]{Ledoux2015}. For comparison with MW and MC measurements we used  calculations with $A_{\rm{V}}^{\rm{max}}=2.0$. Cosmic-ray ionization rate is assumed to be fixed and equal to $10^{-16}$\,s$^{-1}$ that has been estimated for diffuse clouds in Milky-Way \citep{Indriolo2007}. The input parameters for the PDR code are listed in Table\,\ref{PDR_param}. For each model we performed 25 iterations to obtain a stable solution of a cloud structure. Additionally, a convergence of the calculations for each model was checked visually.    

\begin{table*}
\caption{List of input parameters for the PDR Meudon models. The first three parameters ($n_{\rm H}$, $Z$,\UV) were varied within indicated ranges}
\label{PDR_param}
\begin{tabular}{|l|l|c|l|}
\hline
Parameter  & Units & Value & Comment\\
 \hline \hline
 $n_{\rm H}$ & cm$^{-3}$ & $1-10^4$ & Total hydrogen number density\\
 $Z$ & solar & $0.1-3$ & Metallicity \\
 $I_{\rm UV}$ & Mathis field & $0.1-1000$ & UV radiation strength\\
 $I_{\rm CR}$ & s$^{-1}$& $10^{-16}$ & Cosmic ray ionization rate\\
 $A_{\rm Vmax}$ & & 0.5-2 & Maximum visual extinction\\
 $v_{\rm turb}$ & km\,s$^{-1}$ & 2 & Gas turbulent velocity\\
 $T_{\rm CMB}$ & K & 2.73, 9.56 & CMB temperature \\
 ${\rm Ext}$ & & Galaxy & Type of extinction curve\\
 $R_{\rm V}$ & &3.1 & Ratio of the visual extinction to color excess\\
 $C_{\rm D}$ & cm$^{-2}$ & $5.8\times10^{21}/Z$ & Ratio of the column density of neutral gas to visual extinction\\
 $m_{\rm gr}$ & & $0.01\times Z$ & Dust to gas mass ratio\\
 $q_{\rm PAH}$ & & 4.6$\times$10$^{-2}$ & Polycyclic aromatic hydrocarbon fraction\\
 $\alpha_{\rm gr}$ & & 3.5 & Grain power-law distribution index\\
 $r_{\rm min}$ & cm & $10^{-7}$ & Grains minimum radius\\ 
 $r_{\rm max}$ & cm & $3\times10^{-5}$ & Grains maximum radius\\
 \hline
\end{tabular}
\end{table*} 

\begin{figure*}
\begin{center}
        \includegraphics[width=\textwidth]{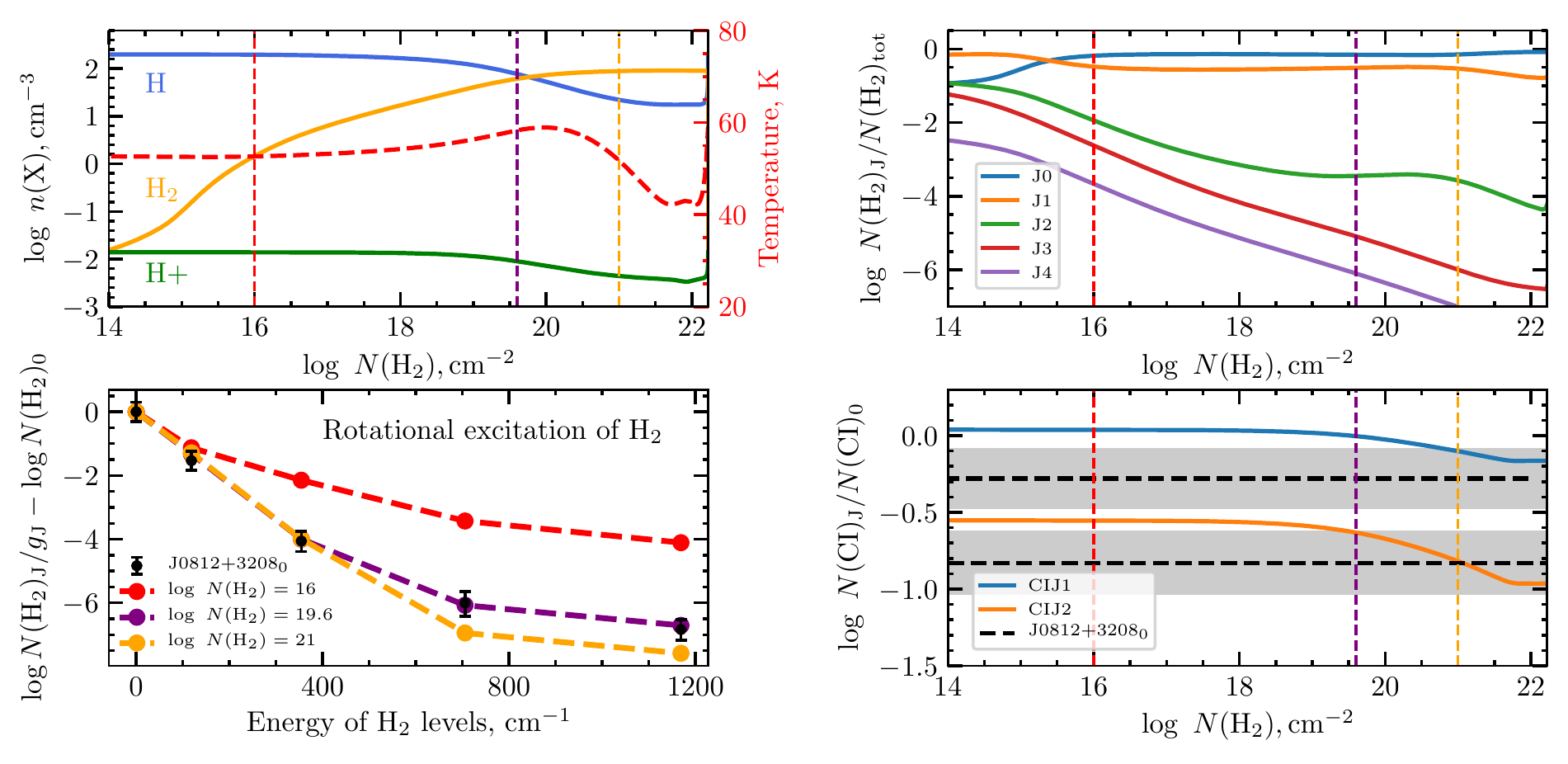}
        \caption{\rm An example of molecular cloud model with the Meudon PDR code. The top left panel presents the number densities of H, H$_2$ and H$^+$ and the temperature (with red dashed line) as a function of the total H$_2$ column density. In the top and bottom right panels we show the profiles of the relative populations of H$_2$ rotational levels and \CI fine structure levels, respectively.
        The bottom left panel shows the comparison of H$_2$ rotational level excitation  calculated for three threshold depths, $\log N({\rm H_2})$ = 16, 19.6 and 21, which are shown by vertical dashed lines in the right panels. The column densities are normalized using the statistical weight of H$_2$ rotational levels, $g(J)$. 
        For comparison, the observed excitation of H$_2$ and \CI\ levels in the absorption system at $z=2.626443$ towards QSO J\,0812$+$3208 are shown by black circles and gray stripes in the bottom left and bottom right panels, respectively.  
        }
        \label{fig01}
\end{center}
\end{figure*}

An example of a PDR model with $Z=0.1$, $n_{\rm{H}}=2\times10^2\,\mbox{cm}^{-3}$ and $I_{\rm UV} = 0.7$ is shown in Fig.\,\ref{fig01}. 
One can see that population of H$_2$ rotational levels in this model are well agree with observed ones (shown in the bottom left panel of Fig.~\ref{fig01}) in the absorption system at $z=2.626443$ towards QSO J\,0812$+$3208 (\citealt{Balashev2010}, J\,0812$+$3208\,A) at the size of the cloud corresponding to the total H$_2$ column density $\log N({\rm H_2})=19.9$, which is close to the measured one. At this depth the relative populations of \CI\ fine-structure levels, are slightly higher than measured in the J\,0812$+$3208\,A (shown in the bottom right panel of Fig.~\ref{fig01}). One can also note that at the mild H$_2$ column densities, $\log N({\rm H_2}) \lesssim 19$ (typical for studied systems), excitation of \CI\ fine-structure levels are more or less constant within the cloud. The drop in the population ratios of \CI\ in the cloud interior is attributed with the temperature drop (see the left top panel of Fig.~\ref{fig01}).

We ran grids of slab constant-density models in which we varied three parameters - the metallicity ($Z$, assuming dust scales to metallicity), total hydrogen number density (\nH) and intensity of UV radiation (\UV). The kinetic temperature, chemical composition, and excitation of the levels were calculated self-consistently by PDR Meudon. To facilitate the representation of the results, for each metallicity in a set [-1.0, -0.5, 0.0, 0.5], we calculate grid of models that uniformly covers the parameter space on a logarithmic scale in the ranges of $0 \leq \log \rm n_{H} / cm^{-3} \leq 4$ (with 10 points) and $-1 \leq \log I_{\rm UV} \leq 3$ (with 9 points).

To compare observed populations of H$_2$ and \CI\ levels in some absorption system with PDR models we firstly choose an appropriate $I_{\rm UV}-n_{\rm H}$ grid with metallicity closest to the observed one. 
Then within chosen grid we calculated the likelihood function based on least-square comparison of observed and model population of levels. To obtain the model population of H$_2$ and \CI\ levels we determine a depth, where the total measured H$_2$ column density, calculated from one side of the model, equals to half the observed one. 
Then we calculated column densities of H$_2$ rotational levels at determined depth, and double the results. This approach mimics a cloud illuminated on both sides with ISRF scaled by $I_{\rm UV}$ factor and with the total column density of H$_2$ equal to the measured one. In order to obtain a smooth representation of the results in $n_{\rm{H}}-I_{\rm{UV}}$ plane, we use an interpolation of the populations in the calculated grid to the denser grid.

For comparison of H$_2$ rotational levels calculated by PDR code with the observed ones we added a factor of 2 (0.3~dex) systematic uncertainty to the observed column densities. This was done to relax the choice of an slab geometry in the modelling. Indeed the UV pumping of H$_2$ strongly depends on the saturation of Lyman and Werner lines, which depends in turn on the H$_2$ column densities. Therefore the real geometry and its deviation from the simplistic shape typically used in modelling (e.g. slab or spherical) can induce a systematic uncertainty, which we tried to account for with this simplistic 0.3\,dex factor. 
A similar systematic was found by \cite{Sternberg2014}, where authors compared various types of geometry (slab, complex, sphere) and found that the difference of the molecular fraction along the line of sight is not higher than 40\% for models with the same H\,{\sc i} optical depth. This difference is accounted to various structure of H\,{\sc i}-H$_2$ transition region, from which the major part of the observed high H$_2$ rotational levels are originated. 
Therefore, following \cite{Sternberg2014} we assumed that a factor of 2 is a good conservative choice to take into account the systematic uncertainty concerned with an unknown geometry of the cloud.  

For \CI\ we compared the relative population of fine structure levels. We use the relative populations, since the absolute abundance of \CI\ is usually hard to reproduce in modelling, since it is determined by the ionization balance of carbon atoms in the ISM (e.g. \citealt{Jenkins2011}), and therefore it depends on many other parameters, such as the ionization fraction, dust properties and  carbon abundance, which are usually poorly constrained in observations. 
However, for studied absorption systems the majority of the UV lines, at which excitation of \CI\ fine-structure levels takes place are usually optically thin.
Therefore the UV excitation of \CI\ is little dependent on total abundance of \CI\ (i.e. the location and \CI/\ion{C}{ii} transition in the cloud) and hence it can be used to constrain UV flux. Therefore we argue that the relative populations of \CI\ fine-structure levels is a good tracer of the physical conditions (UV flux, number density and temperature) inside the cloud. 

\begin{figure*}
\begin{center}
        \includegraphics[width=\textwidth]{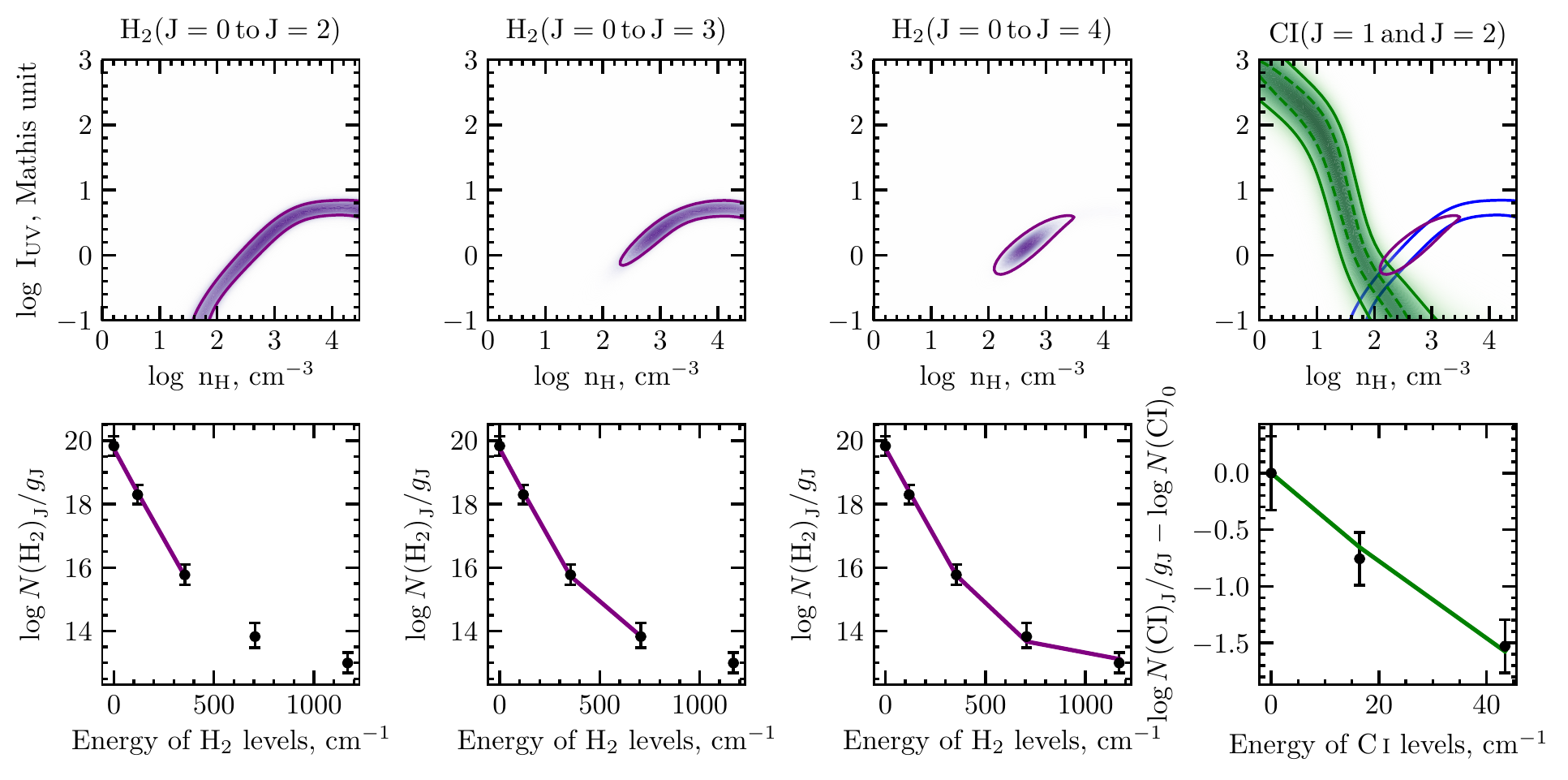}
        \caption{\rm An example of the 
        the analysis of excitation of H$_2$ rotational and \CI\ fine structure levels
        in the absorption system J\,0812$+$3208\,A. The left and central panels correspond to fit to populations of various number of H$_2$ rotational levels, which is marked at the top. The right panels correspond to fit to the relative population of \CI\ fine structure levels.
        In top panels we show the constraints on the number density and UV intensity using excitation of marked levels. 
        The probability density is represented by the colour intensity (purple for H$_2$ and green for C\,{\sc i}), with the 68\%
        confidence level contour shown as the solid 
        line. For \CI\ we additionally show 30\% confidence level by the green dashed line. For best comparison we overplot H$_2$ constraints obtained with fit to (J=0 to J=2) and (J=0 to J=4) levels with the blue and purple contours in the top right panel. In bottom panels we show respectively best fits to the populations of H$_2$ and \CI levels using purple and green  lines. The observed population of H$_2$ levels and \CI\ levels are shown by black dots. Column densities normalized using the statistical weight $g_{\rm J}$ of H$_2$ and \CI\ levels. For \CI\ we normalized populations to the population of the ground C\,{\sc i}(J=0) level as well. 
        }
        \label{fig02}
\end{center}
\end{figure*}

An example of the constraints on the number density and UV intensity using excitation of H$_2$ rotational and \CI\ fine-structure levels is shown in the top panels of Fig.~\ref{fig02} (for the 
J\,0812$+$3208\,A). To show how the various combination of H$_2$ and \CI\ levels constrain the parameter space we plot series of the panels. The best fits to the populations of selected H$_2$ and \CI\ levels are shown in the bottom panels of Fig.~\ref{fig02}. The lower rotational levels of H$_2$ J=0, 1, 2 (the left panel of Fig.~\ref{fig02}) in saturated systems usually correspond to the kinetic temperature in the cloud, and therefore the obtained constraint in $I_{\rm{UV}} - n_{\rm{H}}$ plane reflects the excitation temperature $T_{0-2}$ of H$_2$. In the reasonable range of the number densities corresponded to the cold diffuse medium ($\log n_{\rm H} \sim 2-3$) this translates in almost linear dependence between $ I_{\rm{UV}}$ and $n_{\rm{H}}$ (see the left  column of Fig.\,\ref{fig02}). 
Adding to J=0,1,2 H$_2$ levels the population of higher H$_2$ (J=3 and J=4) rotational levels, results in tighter constraints on the $I_{\rm UV} - n_{\rm H}$ region (see central panels in Fig.\,\ref{fig02}).
This is mainly, since the UV pumping is the main mechanism of excitation of high rotational levels of H$_2$ \citep{Black1976}, therefore their populations are very sensitive to the intensity of UV radiation.  
The constraints on $I_{\rm UV} - n_{\rm H}$ using \CI\ fine-structure is shown in the right panel in Fig.~\ref{fig02}. In most cases \CI\ gives quite wide degenerate region in $I_{\rm UV} - n_{\rm H}$ plane, but it typically is nearly orthogonal to the region constrained using H$_2$ \citep[see also][]{Balashev2019}. Therefore a joint \CI-H$_2$ fit allows us to break the degeneracy and provides significantly tighter constraints. Note that, in a case of J\,0812$+$3208\,A a joint fit is well consistent with the constraint obtained with H$_2$ levels alone.

\subsection{Note on excitation of high H$_2$ rotational levels.}
\begin{figure*}
\begin{center}
        \includegraphics[width=\textwidth]{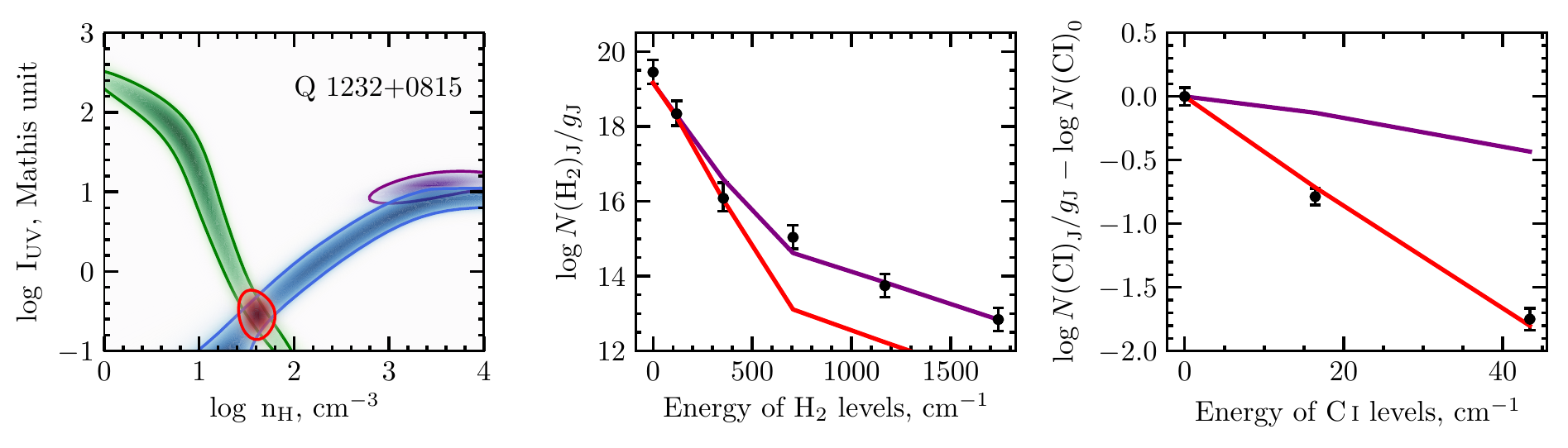}
        \caption{\rm An example of the analysis of excitation of H$_2$ rotational and \CI\ fine structure levels in the H$_2$ absorption system at $z=2.3377$ towards QSO J\,1232$+$0815. In the left panel green, blue and purple contours represent the constraints on the number density and UV intensity obtained with the analysis of excitation of fine structure levels of \CI, lower rotational levels of H$_2$ (J=0 to J=2) and both lower and higher rotational levels of H$_2$ (J=0 to J=5), respectively. The red contour indicate a joint constraint using \CI\ and lower levels of H$_2$. The middle and right panels show the population of H$_2$ rotational and \CI\ fine structure levels, respectively. The black circles indicate the observed values. The purple lines show the model with the best fit values of the number density and UV intensity obtained from the fit to all rotational levels of H$_2$ (J=0 to J=5), while the red lines correspond to the joint fit to population of lower (J=0 to J=2) levels of H$_2$ and \CI\ levels.
        }
        \label{fig04}
\end{center}
\end{figure*}

In some absorption systems we were not able to describe simultaneously and self-consistently the excitation of rotational levels of H$_2$ and fine-structure levels of \CI\ within a one-component PDR model. For illustration, in Fig.\,\ref{fig04} we show an example of such problematic fit for H$_2$-bearing DLA towards QSO J\,1232$+$0815. 
We found that in this system the  population of high rotational levels of H$_2$ (at J=3, 4, 5) is relatively enhanced. In one component model such an excitation gives the UV intensity about 10 times higher than Mathis field. To satisfy the thermal balance and kinetic temperature ($T_{\rm{01}}\sim70\,K$, which is mostly constrained by low rotational levels of H$_2$), a high gas density, $n_{\rm{H}}\sim10^3\mbox{\,cm}^{-3}$, is required. At such physical conditions excited \CI\ levels should be significantly populated, much higher than it is really observed. 



To reconcile this discrepancy more complex model should be invoked. This discrepancy is most likely arisen from the fact that different H$_2$ rotational levels probed the different regions within the cloud. Indeed, one can see in top right panel of Fig.~\ref{fig01} that high rotational levels of H$_2$ are strongly excited in the outer regions of the clouds, at low H$_2$ column densities, since UV pumping lines are not saturated and hence the photoexcitation rate is relatively high. Such "outer regions" of the cloud actually attributed to the diffuse atomic medium, with relatively low H$_2$ molecular fraction. It is likely that hydrodynamical motions and turbulence mixing can increase the size of this region. Indeed, in the static model, the H$_2$ abundance in the cloud is set by the balance between the formation on dust and the photodestruction by UV field, which is accompanying process to the photoexcitation, taken place in the Werner and Lyman UV lines. In the outer part of the cloud the lines are not saturated and therefore the photodestruction rate per one H$_2$ molecule is much higher than the rate of H$_2$ formation on the dust grain. Therefore the outer envelope of the cloud has relatively low H$_2$ molecular fraction $\sim 10^{-5}$. This is opposite in the inner self-shielded part, where photodestruction drastically dropped, due to saturation of the lines, and the hydrogen converts to molecular form. Once we add the hydrodynamical mixing of the layers to the static model, we will get that if a self-shielded part of the medium moves to the non-self-shielded region, the hydrogen will quickly (
on timescales $t_{\rm diss} \sim \left(10^{10} - 10^{11}\right) \times I_{\rm UV}$\,s, correspond to the unshielded photoexcitation at ISM UV field) convert from molecular to dominantly atomic form. Oppositely, if atomic medium will come to the shielded region, it will take a much more time ($t_{\rm form} \sim 10^{15} Z^{-1}$\,s) to convert the hydrogen to molecular form. This qualitatively explain, why we can expect that the diffuse atomic envelope of the cloud can be more extended than ones in static models. As we already show, the high rotational levels of H$_2$ are significantly excited in this region and therefore one may expect enhanced excitation of these levels in comparison to  the static model results. Additionally, high rotational level of H$_2$ can be additionally excited in outer envelopes by a turbulent dissipation, which is expected to be also enhanced in this region (e.g., \citealt{Cecchi2005}). Moreover, there is an observational signature of a such enhancement, concerned with the well established effect of the doppler parameter increase with an increase of the H$_2$ rotational level number \citep{Noterdaeme2007, Balashev2009}, which is naturally explained by the enhanced doppler parameter in the outer shell of the cloud \citep{Balashev2009}. 
Contrary, the central shielded-part of the clouds is more stable for this mixing and the measured column densities of lower rotational levels of H$_2$ mostly probe
the central parts. Additionally, as we already noted, the low-J levels mostly reflect the thermal balance in the cloud, which also is less affected to the UV radiative transfer in H$_2$ lines. Therefore we argue that such levels provide more reliable measurements of the physical conditions. And in the cases where the population of high rotational levels H$_2$ contradict with physical conditions inferred from the joint analysis of lower H$_2$ rotational levels and \CI\ fine-structure levels, we report the values of the $I_{\rm UV}$ and $n$ neglecting population of high rotational levels of H$_2$.

\section{Data sample}
\label{sec:sample}

We compiled samples of absorption systems with measured populations of H$_2$ rotational levels and \CI\ fine-structure levels at high redshift H$_2$-bearing DLAs and along the lines of sight towards stars in the Milky-Way and the Magellanic Clouds galaxies. 

\subsection{High redshift DLAs} 
Currently about 50 H$_2$-bearing DLAs are known at high redshift (e.g. see \citealt{Balashev2019}). We select high-z DLAs, which were detected in quasar spectra obtained with the Ultraviolet and Visual Echelle Spectrograph (UVES, \citealt{Dekker2000}) mounted on the ESO VLT-UT 2 8.2 m telescope on Cerro Paranal, Chile and the High Resolution Echelle Spectrometer (HIRES, \citealt{Vogt1994}) on the Keck I telescope on Maunakea in Hawaii. These spectra have a high spectral resolution ($R\ge50\,000$ or the width of instrument function $\le6\mbox{\,km\,s}^{-1}$) and a high signal to noise ratio ($S/N>10$).  We do not use the data for H$_2$-bearing DLA systems found in spectra obtained with medium spectral resolution (e.g. observed with VLT/X-shooter with $R\sim5000-9000$), while such estimates are possible \cite[see][]{Balashev2019}. This was done, to minimize a possible systematic uncertainty concerned with the determination of a velocity structure. Indeed,  complex multicomponent velocity structure in H$_2$ lines seen in at least half of H$_2$-bearing DLAs,
which is very hard to unambiguously resolve with intermediate resolution spectra. The high resolution spectra provide this opportunity and even allow to reliably measure the Doppler parameters of H$_2$ and \CI\ absorption lines with $\sim2-5\mbox{\,km\,s}^{-1}$, due to large number of lines involved in analysis. 

To compile a final sample we also used two additional criteria: (i) we used only H$_2$-bearing DLAs where \CI\ were detected, since a joint analysis of H$_2$ and \CI\ allows to significantly improve constraints. (ii) We used only the  systems with the total H$_2$ column density higher than $\log N(\rm H_2)\sim 18$. This was done to be sure that we really probed self-shielded regions. Additionally, we found using the closer look in {\sc PDR Meudon} results, that for most systems satisfied these criteria, low rotational levels are thermalized, i.e. $T_{\rm 01}$ close to the kinetic temperature in the medium. 

After applying these criteria, we are left with twelve H$_2$-bearing DLAs at $z>1.9$, containing totally of fourteen velocity components. This sample, that we call $S^{\rm{DLA}}$, is presented in Table\,\ref{list}. 

\begin{table*}
\caption{The list of H$_2$ absorption systems included in $S^{\rm{DLA}}$ sample. The columns are: (1) name of QSO, (2) the redshifts of DLA, (3) H\,{\sc i} column densities, (4) total H$_2$ column density (5) total \CI\ column density, (6) an average metallicity, (7) $T_{\rm 01}$  excitation temperature of H$_2$ (7), our estimates of the hydrogen number density (8) and intensity of incident UV radiation (9).}
\label{list}
\begin{tabular}{|l|l|c|c|c|c|c|c|c|c|}
\hline
Name & ~~$z_{\rm abs}$~~ & ~$\log N_{\rm H\,I}^{\rm DLA}$~ & ~$\log N_{\rm H_2}$~ & 
~$\log N_{\rm CI}$~ & ${\rm[X/H]}$ & $T_{\rm 01}$ & $\log n_{\rm H}$ & $\log I_{\rm UV}$ & Ref\\
 &  & $[\mbox{cm}^{-2}]$ & $[\mbox{cm}^{-2}]$ & $[\mbox{cm}^{-2}]$ &  & [K] & $[\mbox{cm}^{-3}]$ & [Mathis unit] & \\
\hline 
J\,0000$+$0048 & 2.525458 & 20.80$\pm$0.10 & 20.44$\pm$0.03 & 16.21$\pm$0.07 & 0.46$\pm$0.45 & 52$\pm$2 & $1.31^{+0.24}_{-0.42}$ & $-0.19^{+0.28}_{-0.24}$& 1\\
B\,0528$-$2505 & 2.810995 & 21.35$\pm$0.07 & 18.10$\pm$0.01 &  12.15$^{+0.07}_{-0.05}$ & -0.91$\pm$0.07 & $141^{+6}_{-6}$ & $2.22^{+0.21}_{-0.22}$ & $0.90^{+0.14}_{-0.15}$& 2\\
J\,0812$+$3208 & 2.626443 & 21.35$\pm$0.10 & 19.93$\pm$0.05 &   13.52$\pm$0.15 & -0.81$\pm$0.10 & 48$\pm$2 &  $2.26^{+0.23}_{-0.21}$ & $-0.14^{+0.17}_{-0.17}$& 3 \\
              & 2.626276 &                & 18.82$\pm$0.37 &   12.85$\pm$0.02 & -0.81$\pm$0.10 & $50^{+44}_{-16}$ & $0.86^{+0.24}_{-0.42}$ & $-0.95^{+0.34}_{-0.05}$ &  3 \\
J\,0816$+$1446 & 3.28742  & 22.00$\pm$0.10  & $18.62^{+0.21}_{-0.17}$ &   13.67$\pm$0.02 & -1.10$\pm$0.10 & $110^{+33}_{-43}$ & $1.63^{+0.08}_{-0.12}$ & $-0.32^{+0.20}_{-0.16}$& 4 \\
J\,0843$+$0221 & 2.786459 & 21.82$\pm$0.11 & 21.21$\pm$0.02 &   13.52$\pm$0.05 &  -1.52$\pm$0.10 & $123^{+9}_{-8}$ & $1.85^{+0.07}_{-0.09}$ & $1.80^{+0.15}_{-0.14}$& 5 \\
                     & 2.786582 &                & 21.21$\pm$0.02 &   13.79$\pm$0.05 & -1.52$\pm$0.10 & $123^{+9}_{-8}$ & $1.94^{+0.08}_{-0.07}$ & $1.80^{+0.12}_{-0.09}$& 5 \\
J\,1232$+$0815 & 2.3377 & 20.90$\pm$0.08 & 19.57$\pm$0.10  &  14.07$\pm$0.05 & -1.35$\pm$0.12 & $66^{+19}_{-12}$ & $1.58^{+0.14}_{-0.11}$ & $-0.55^{+0.21}_{-0.20}$ & 6 \\
J\,1237$+$0647                & 2.68955 & 20.00$\pm$0.15     & $19.20^{+0.13}_{-0.12} $  &  13.08$\pm$0.02 & 0.34$\pm$0.12 & $108^{+92}_{-34}$ & $1.27^{+0.14}_{-0.10}$ & $0.99^{+0.11}_{-0.15}$& 7 \\
J\,1439$+$1118 & 2.41837 & 20.10$\pm$0.10 & 19.52$\pm$0.07 &  14.64$\pm$0.03 &  0.16$\pm$0.11 & $107^{+33}_{-20}$ & $0.90^{+0.15}_{-0.18}$ & $0.63^{+0.19}_{-0.17}$ & 8\\
B\,1444$+$0126 & 2.08696 & 20.25$\pm$0.07 & 18.15$\pm$0.10  & 13.18$\pm$0.10 & -0.80$\pm$0.09 & $193^{+282}_{-70}$ & $1.94^{+0.26}_{-0.21}$ & $0.49^{+0.16}_{-0.19}$ & 9\\
J\,1513$+$0352 & 2.463622 & 21.82$\pm$0.02 & 21.31$\pm$0.01  & 15.02$\pm$0.05 & -1.22$\pm$0.10 & $82^{+4}_{-4}$ & $1.90^{+0.13}_{-0.12}$ & $0.45^{+0.24}_{-0.20}$ & 10 \\
J\,2100$-$0641 & 3.09145 & 21.05$\pm$0.15 & 18.76$\pm$0.03 &  12.77$\pm$0.03 & -0.73$\pm$0.15 & $159^{+44}_{-29}$ & $1.40^{+0.28}_{-0.35}$ & $-0.50^{+0.22}_{-0.34}$ & 11\\
J\,2140$-$0321 & 2.3399 & 22.40$\pm$0.10 & 20.13$\pm$0.07 & 13.57$\pm$0.03 & -1.05$\pm$0.12 & $75^{+12}_{-9}$ & $2.42^{+0.11}_{-0.09}$ & $1.64^{+0.19}_{-0.21}$ & 12\\
\hline
\end{tabular}
\begin{tablenotes}
\item {\bf References:} (1) \cite{Noterdaeme2017};  (2) \cite{Klimenko2015b}; (3) \cite{Balashev2010}; (4) \cite{Guimaraes2012};
(5) \cite{Balashev2017}; (6) \cite{Balashev2011}; (7) \cite{Noterdaeme2010}; (8) \cite{Srianand2008}; (9) \cite{Ledoux2003}; (10) \cite{Ranjan2018}; (11) \cite{Balashev2015}; (12) \cite{Noterdaeme2015}.
\end{tablenotes}
\end{table*}

\begin{table*}
\caption{List of H$_2$ absorption systems included in $S^{\rm{MW}}$ and $S^{\rm{MC}}$ samples. The columns are the same as in Table~\ref{list}, except the redshift column that is not provided, since these measurements are from local galaxies.}
\label{listMW}
\begin{tabular}{|l|c|c|c|c|c|c|c|c|}
\hline
Name &  ~$\log N_{\rm H\,I}$~ & ~$\log N_{\rm H_2}$~ & 
~$\log N_{\rm CI}$~ & ${\rm[X/H]}$ & $T_{\rm 01}$ & $\log n_{\rm H}$ & $\log I_{\rm UV}$ & Ref\\
  & $[\mbox{cm}^{-2}]$ & $[\mbox{cm}^{-2}]$ & $[\mbox{cm}^{-2}]$ &  & [K] & $[\mbox{cm}^{-3}]$ & [Mathis unit] & \\
\hline
\multicolumn{9}{c}{Milky-Way}\\
HD\,24534 & 20.73$\pm$0.06 & 20.92$\pm$0.03 & 13.97$\pm$0.05 & $-0.2$ & $57^{+3}_{-3}$ & $2.62^{+0.10}_{-0.06}$ & $-0.05^{+0.11}_{-0.11}$ & 1,2 \\
HD\,27778 & 20.98$\pm$0.30 & 20.79$\pm$0.03 & 15.08$\pm$0.05 & $-0.2$ & $56^{+5}_{-5}$ & $2.04^{+0.13}_{-0.09}$ & $-0.23^{+0.20}_{-0.17}$ &  1,2\\
HD\,40893  &  ..          & 20.58$\pm$0.03 & 14.95$\pm$0.05 & $-0.2$ & $78^{+9}_{-8}$ & $1.76^{0.09}_{-0.08}$ & $-0.59^{+0.22}_{-0.16}$ & \\
HD\,147888  & ..      & 20.48$\pm$0.03 & 14.23$\pm$0.05 & $-0.2$ & $45^{+3}_{-3}$ & $2.71^{+0.07}_{-0.09}$ & $0.09^{+0.17}_{-0.13}$ &  1,2\\
HD\,185418  & 21.11$\pm$0.15 & 20.77$\pm$0.03 & 14.82$\pm$0.05 & $-0.2$ & $101^{+10}_{-8}$ & $1.81^{+0.06}_{-0.08}$ & $-0.32^{+0.25}_{-0.17}$ & 1,2 \\
HD\,192639  & 21.32$\pm$0.12 & 20.69$\pm$0.03 & 14.99$\pm$0.05 & $-0.2$ & $98^{+9}_{-9}$ & $1.94^{+0.08}_{-0.04}$ & $-0.14^{+0.22}_{-0.21}$ &  1,2\\
HD\,195965  & .. & 20.37$\pm$0.03 & 14.67$\pm$0.05 & $-0.2$ & $110^{+7}_{-7}$ & $1.85^{+0.12}_{-0.09}$ & $-0.32^{+0.22}_{-0.22}$ & \\
HD\,206267  & 21.30$\pm$0.15 & 20.86$\pm$0.03 & 15.54$\pm$0.05 & $-0.2$ & $64^{+3}_{-3}$ & $2.13^{+0.06}_{-0.08}$ & $-0.32^{+0.17}_{-0.13}$ &  1,2\\
HD\,207198  & 21.34$\pm$0.17 & 20.83$\pm$0.03 & 15.53$\pm$0.05 & $-0.2$ & $66^{+3}_{-3}$ & $2.04^{+0.05}_{-0.76}$ & $-0.28^{+0.18}_{-0.23}$ &  1,2\\
HD\,210839  & 21.15$\pm$0.10 & 20.84$\pm$0.03 & 14.41$\pm$0.05 & $-0.2$ & $72^{+4}_{-4}$ & $3.30^{+0.15}_{-0.10}$ & $0.54^{+0.12}_{-0.11}$ &  1,2\\
 &  &  20.84$\pm$0.03 & 14.99$\pm$0.05 & $-0.2$ & $72^{+4}_{-4}$ & $2.04^{+0.09}_{-0.05}$ & $-0.32^{+0.20}_{-0.13}$ & \\
\hline 
\multicolumn{9}{c}{LMC}\\
SK\,$-67^{\circ}05$ & 20.88$^{+0.12}_{-0.15}$ & 19.46$\pm$0.05 & 13.62$\pm$0.02 & $-0.4$ & $57^{+5}_{-4}$ & $2.22^{+0.13}_{-0.10}$ & $-0.68^{+0.16}_{-0.20}$ &  3,4\\ 
&  & 19.46$\pm$0.05 & 13.15$\pm$0.04 & $-0.4$ & $57^{+5}_{-4}$ & $2.13^{+0.19}_{-0.19}$ & $-0.86^{+0.20}_{-0.14}$ &  \\ 
SK\,$-70^{\circ}115$&  21.37$^{+0.12}_{-0.15}$ & 19.94$\pm$0.07 & 13.25$\pm$0.04 & $-0.4$ & $53^{+6}_{-5}$ & $1.63^{+0.18}_{-0.17}$ & $-0.37^{+0.35}_{-0.31}$ &  3,4\\ 
 & & 19.94$\pm$0.07 & 13.35$\pm$0.04 & $-0.4$ & $53^{+6}_{-5}$ & $1.81^{+0.12}_{-0.13}$ & $-0.28^{+0.29}_{-0.27}$ &  \\ 
 & & 19.94$\pm$0.07 & 13.70$\pm$0.03 & $-0.4$ & $53^{+6}_{-5}$ & $2.31^{+0.16}_{-0.15}$ & $0.04^{+0.26}_{-0.20}$ &  \\ 
\hline 
\multicolumn{9}{c}{SMC}\\
SK 13 & 21.15$\pm$0.10 & 20.36$\pm$0.07 & 13.12$\pm$0.04 & $-0.7$ & $66^{+5}_{-5}$ & $2.08^{+0.14}_{-0.17}$ & $-0.23^{+0.26}_{-0.20}$ & 4,5,6\\
 &  & 20.36$\pm$0.07 & 13.51$\pm$0.09 & $-0.7$ & $66^{+5}_{-5}$ & $2.67^{+0.30}_{-0.24}$ & $0.18^{+0.24}_{-0.23}$ & \\
 &  & 20.36$\pm$0.07 & 13.54$\pm$0.05 & $-0.7$ & $66^{+5}_{-5}$ & $3.12^{+0.25}_{-0.49}$ & $0.18^{+0.10}_{-1.18}$ & \\
  &  & 20.36$\pm$0.07 & 13.31$\pm$0.06 & $-0.7$ & $66^{+5}_{-5}$ & $2.94^{+0.31}_{-0.26}$ & $0.31^{+0.24}_{-0.21}$ & \\
SK 18 & 21.90$\pm$0.15 & 20.63$\pm$0.05  & 13.13$\pm$0.06 & $-0.7$ & $53^{+4}_{-4}$ & $1.94^{+0.16}_{-0.16}$ & $-0.23^{+0.25}_{-0.36}$ & 3,4,5\\
 & & 20.63$\pm$0.05 & 14.04$\pm$0.03 & $-0.7$ & $53^{+4}_{-4}$ & $2.76^{+0.13}_{-0.10}$ & $0.27^{+0.16}_{-0.27}$ & \\
\hline
\end{tabular}
\begin{tablenotes}
\item {\bf References:} (1) \cite{Jensen2010}; (2) \cite{Jenkins2011}; (3) \cite{Tumlinson2002}; (4) \cite{Welty2016}; (5) \cite{Rachford2002}; (6) \cite{Cartledge2005}
\item Values of the metallicity are taken from \cite{Martinez2002} for the Milky-Way and references in \cite{Welty2016} for the LMC and SMC.
\end{tablenotes}
\end{table*} 

\subsection{Milky-Way and the Magellanic Clouds}

Absorption line measurements of H$_2$ and \CI\ in the ISM of galaxies in the local Universe (the Milky Way and the Small and Large Magellanic Clouds) were performed with the Far Ultraviolet Spectroscopic Explorer (FUSE) and the Space Telescope Imaging Spectrograph (STIS) on board the Hubble Space Telescope (HST). The FUSE spectra have the moderate spectral resolution 
(approximately $R = 20000$ or ${\rm{FWHM}}\sim15\mbox{\,km\,s}^{-1}$) and cover the range $905-1180$\,\AA\,. These spectra were used to analyse H$_2$ absorption systems in the Milky-Way by \cite{Snow2000, Rachford2002, Rachford2009, Jensen2010} and Magellanic Clouds by \cite{Tumlinson2002}. The STIS spectra have the high spectral resolution up to $R = 114000$ or ${\rm{FWHM}}=2.7\mbox{\,km\,s}^{-1}$ and cover the range $\lambda=1160-3100$\,\AA\, in the E140H and E230H modes. The spectra obtained by STIS were used to analyse \CI\ absorptions in MW by \cite{Jenkins2011}, and LMC and SMC by \cite{Welty2016}.

We selected ten and four H$_2$ absorption systems observed in the Milky-Way and Magellanic Clouds, respectively, in which populations of both H$_2$ rotational levels and \CI\ fine-structure levels were measured. 
These samples, that we call $S^{\rm{MW}}$ and $S^{\rm{MC}}$, are presented in Table\,\ref{listMW}.
Each absorption system in the $S^{\rm MC}$ sample has several velocity components. These components are resolved in \CI\ lines in STIS spectra, but not resolved in H$_2$ lines since FUSE spectra have lower spectral resolution and H$_2$ lines from $J<3$ levels are highly saturated ($\log N({\rm H_2})\sim19-20$ for J=0,1). We assume that individual \CI\ components are embedded in one huge cloud H$_2$, excitation of which is described by some average kinetic temperature. The variation in the kinetic temperature between the components is probably offset by a systematic uncertainty ($\pm$0.3~dex), which we added to the observed population of H$_2$ levels. Therefore we analysed individually each \CI\ component, but used the total rotational populations of H$_2$ levels.

\section{Results}
\label{sec:results}

In this section we present the constraints on the $T_{\rm{kin}}$, $I_{\rm UV}$ and $n_{\rm H}$ obtained by an analysis of H$_2$ and \CI\ excitation in high-redshift DLAs and local galaxies (Milky-Way, LMC and SMC) and discuss the obtained results.
Our estimates are presented in Table\,\ref{list} for the $S^{\rm{DLA}}$ sample, and in Table\,\ref{listMW} for the $S^{\rm{MW}}$ and  $S^{\rm{MC}}$ samples. 
The figures contain fit to H$_2$ and \CI\ population, and 2D probability density functions for $I_{\rm UV}-n$, for each systems in considered samples are presented in Appendix\,\ref{ApppxA}.
\begin{figure*}
\begin{center}
        \includegraphics[width=\textwidth]{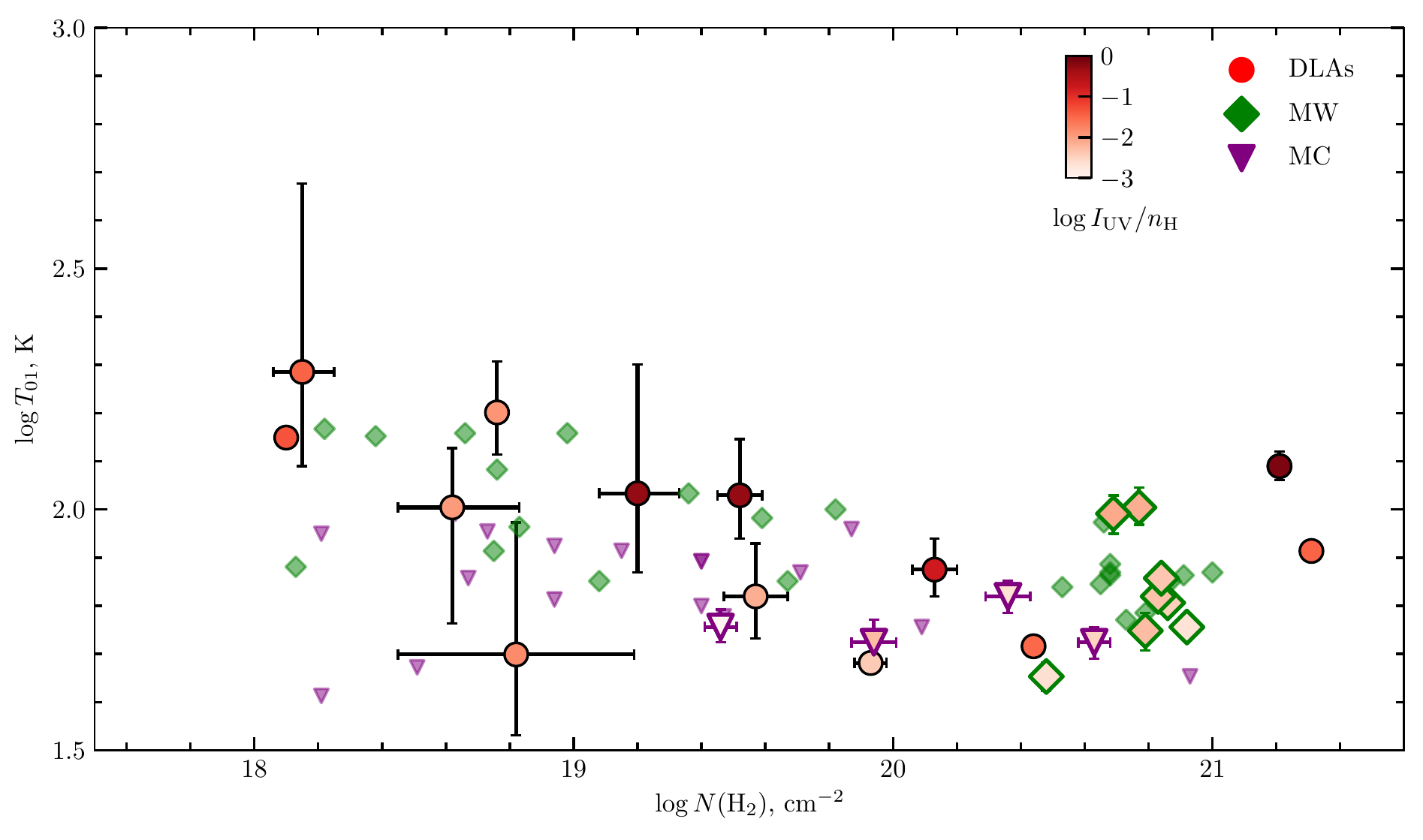}
        \caption{The excitation temperatures $T_{\rm{01}}{\rm(H_2)}$ as a function of the column density of H$_2$ systems measured in high redshift DLAs (circles), Milky-Way (diamonds) and Magellanic Clouds (triangles). Color gradient of points encodes our estimate of the ratio of the intensity of incident UV field to number density. Little green diamonds and purple triangles respectively represent known data of $T_{\rm{01}}$ in H$_2$ systems in the Milky-Way \citep{Gillmon2006}, SMC and LMC \citep{Tumlinson2002}.}
        \label{nuvplane}
\end{center}
\end{figure*}

\subsection{Kinetic temperature}
\label{sec:T01}

We used $T_{\rm 01}$ excitation temperature as a measurement of the kinetic temperature.
The $T_{\rm 01}$ excitation temperature as a function of $\log N({\rm H_2})$ is shown in Fig.\,\ref{nuvplane}.
We found that $T_{\rm 01}$ in the $S^{\rm{DLA}}$ sample is in the range of $47-190\mbox{\,K}$ with the mean about $\sim100$\,K. 
In the $S^{\rm{MW}}$ we found slightly lower temperatures in range 
($44-100\mbox{\,K}$). This is not surprisingly, since $S^{\rm MW}$ sample probes higher H$_2$ column densities, due to only such systems 
have been studied in the HST/STIS \CI\ survey. For these column densities, well-established trend (which was reported for both local and high-z measurements \cite{Muzahid2015, Balashev2017}) of a decrease in the $T_{\rm{01}}$ temperatures with an increase in the H$_2$ column density,  suggests typically lower temperatures in $S^{\rm MW}$. To illustrate this trend we additionally plot by little green diamonds the measurements of $T_{\rm 01}$ in H$_2$ systems in the Milky-Way, where \CI\ absorptions were not detected \citep{Gillmon2006, Jensen2010}. 
Based on the detailed look of the PDR models we found that this trend can be explained by an enhanced H$_2$ molecular fraction inside the cloud with an increase of the H$_2$ column density. The enhanced H$_2$ molecular fraction increases the cooling rate by H$_2$ \citep[e.g.][]{LePetit2006}. For $\log N (\rm H_2) \gtrsim 19$ cooling by H$_2$ begins to play a significant role in the thermal balance in comparison with cooling by \CII\ and \OI\ lines which dominate in the atomic medium.

\begin{figure*}
\begin{center}
        \includegraphics[width=\textwidth]{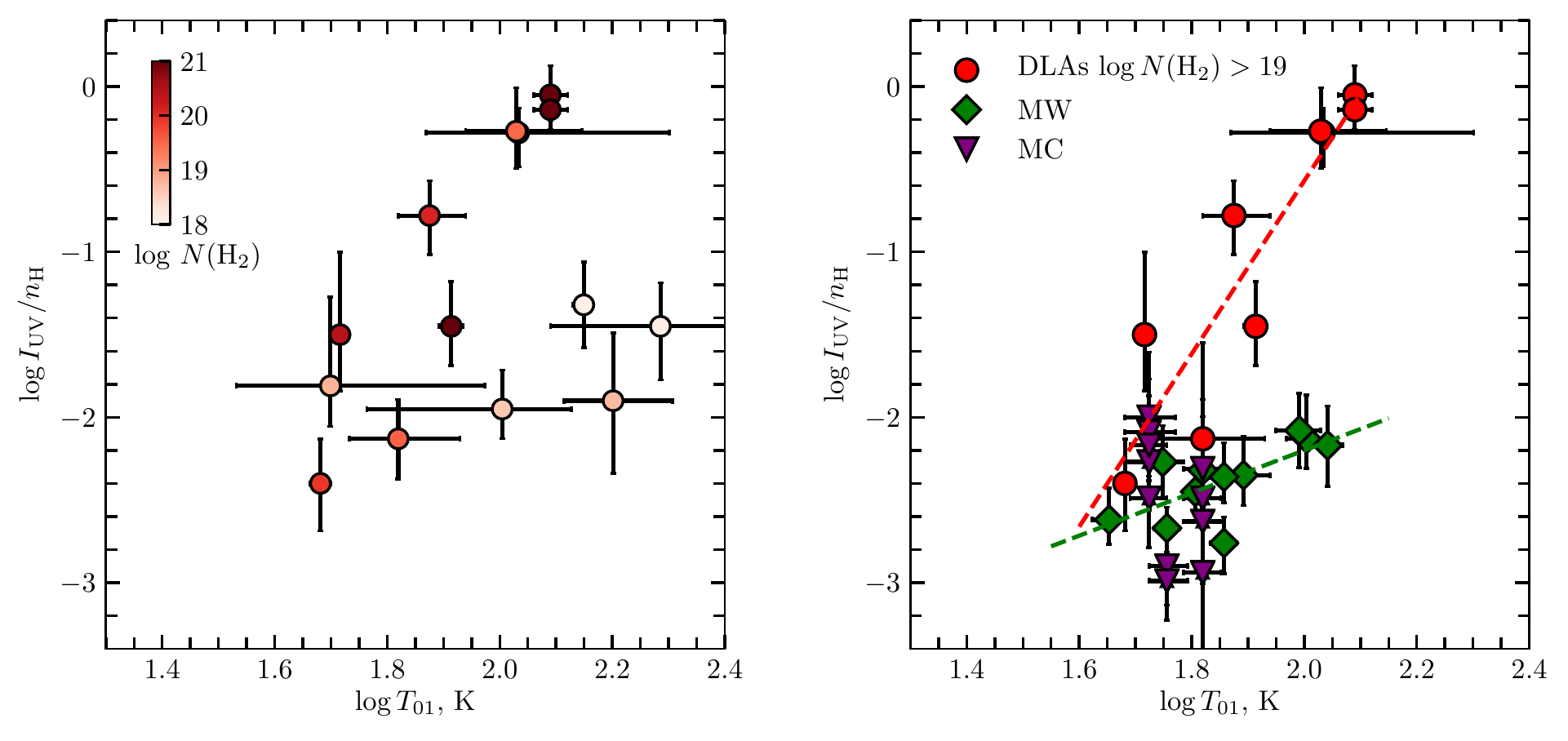}
        \caption{\rm The ratio of the UV intensity to number density, $I_{\rm UV}/n_{\rm H}$, versus the excitation temperature $T_{\rm01}$. The $I_{\rm UV}$ is in the Mathis units and $n_{\rm H}$ in cm$^{-3}$. In the left panel we show data for high redshift H$_2$-bearing DLAs. Color gradient encodes the H$_2$ column density. In the right panel we compare measurements, obtained in only saturated H$_2$ systems with $\log N>19$ in high redshift DLAs (red circles), Milky-Way (green diamonds) and Magellanic Clouds (purple triangles). The dashed lines represent our fit to MW and DLA samples.}
        \label{T01_fig}
\end{center}
\end{figure*}

In the $S^{\rm MC}$ sample we get $T_{\rm 01}$ within the range of $52-64\mbox{\,K}$, which is systematically lower than ones in the $S^{\rm DLA}$ and $S^{\rm MW}$ samples at the similar column densities (taking into account aforementioned trend). This difference is much probably attributed to the difference in the physical conditions. 
Indeed, we show this with a red color gradient of symbols in Fig.\,\ref{nuvplane}. The color encodes the ratio of UV radiation intensity to number density  ($I_{\rm{UV}}/n_{\rm{H}}$). One can note that at specified H$_2$ column density the kinetic temperature  increases with an increase of this ratio. This is better seen if we plot the ratio of $I_{\rm UV}/n_{\rm H}$ as a function of $T_{\rm 01}$, see Fig.\,\ref{T01_fig}.  
In the left panel we plot this relation for the $S^{\rm DLA}$ sample. One can see (the right panel of Fig.\,\ref{T01_fig}) that for a subsample, $S^{\rm{DLA}}_{19}$, of H$_2$ absorption systems with $\log N({\rm H_2})>19$
there is an evident power law dependence over the range of 
$1.5\lesssim\log T\lesssim2.2$ with the index $\alpha_{\rm DLA}=5.2\pm0.8$. For MW sample the power law dependence is less evident, while the data formally suggests index $\alpha_{\rm MW}=1.3\pm0.2$.  

Such power law dependence is naturally expected from the thermal balance that is implemented in the {\sc PDR Meudon} code. Indeed, derived estimates on $I_{\rm UV}$ and $n_{\rm H}$ are based on the observed H$_2$ orto-para ratio which for saturated H$_2$ systems, considered here matches with kinetic temperature. The thermal balance suggests, that in diffuse ISM the cooling is dominated by \ion{C}{II} emission and heating is mostly determined by the photoelectric heating. Roughly one can write (see e.g. eqs.\,(19) and (20) from \citealt{Wolfire2003})
\begin{equation}
\frac{I_{\rm{UV}}}{n_{\rm H}} \propto T^{\alpha},  
\label{eq_th_balance}
\end{equation}
where $\alpha$ is the index, which actually depends on the dust properties, and physical parameters, e.g. the standard combination of $I_{\rm UV} T^{1/2} / n_e$, where $n_e$ is the electron density \citep[see][]{Wolfire2003}. Since  metallicities (and hence electron fractions) are different in the DLA and MW samples, this can explain the difference seen in the power law indexes. Also note that MW sample is located in very tight range of $I_{\rm UV} / n_{\rm H}$, while DLA sample spans over three orders of magnitudes.  

\subsection{Number density}

Derived hydrogen number densities in $S^{\rm{DLA}}$ sample are in the range $10-300\mbox{\,cm}^{-3}$ with the average about $50\mbox{\,cm}^{-3}$. This is about $\sim5$ times lower than average values of $\sim200\mbox{\,cm}^{-3}$ derived in both $S^{\rm MW}$ and $S^{\rm MC}$ samples. The most evident explanation of this difference is, that these samples probe different column densities ranges. Indeed, the MW sample corresponds to the higher end of the H$_2$ column density distribution, i.e. probes deeper into the cloud, where higher number densities are naturally expected. Additionally, the DLA sample probes the remote galaxies, that are predominantly "blindly" selected , i.e. selected by a cross-section. Such a technique evidently probes preferentially H$_2$-bearing gas, located in outskirts of the remote galaxies, Therefore,  it is reasonable to expect that absorption systems in the DLA sample corresponds to less dense gas than the gas in the MW sample, which is observed in $\sim$ kpc vicinity around the Sun.

\subsection{Thermal pressure}

\begin{figure*}
\begin{center}
        \includegraphics[width=\textwidth]{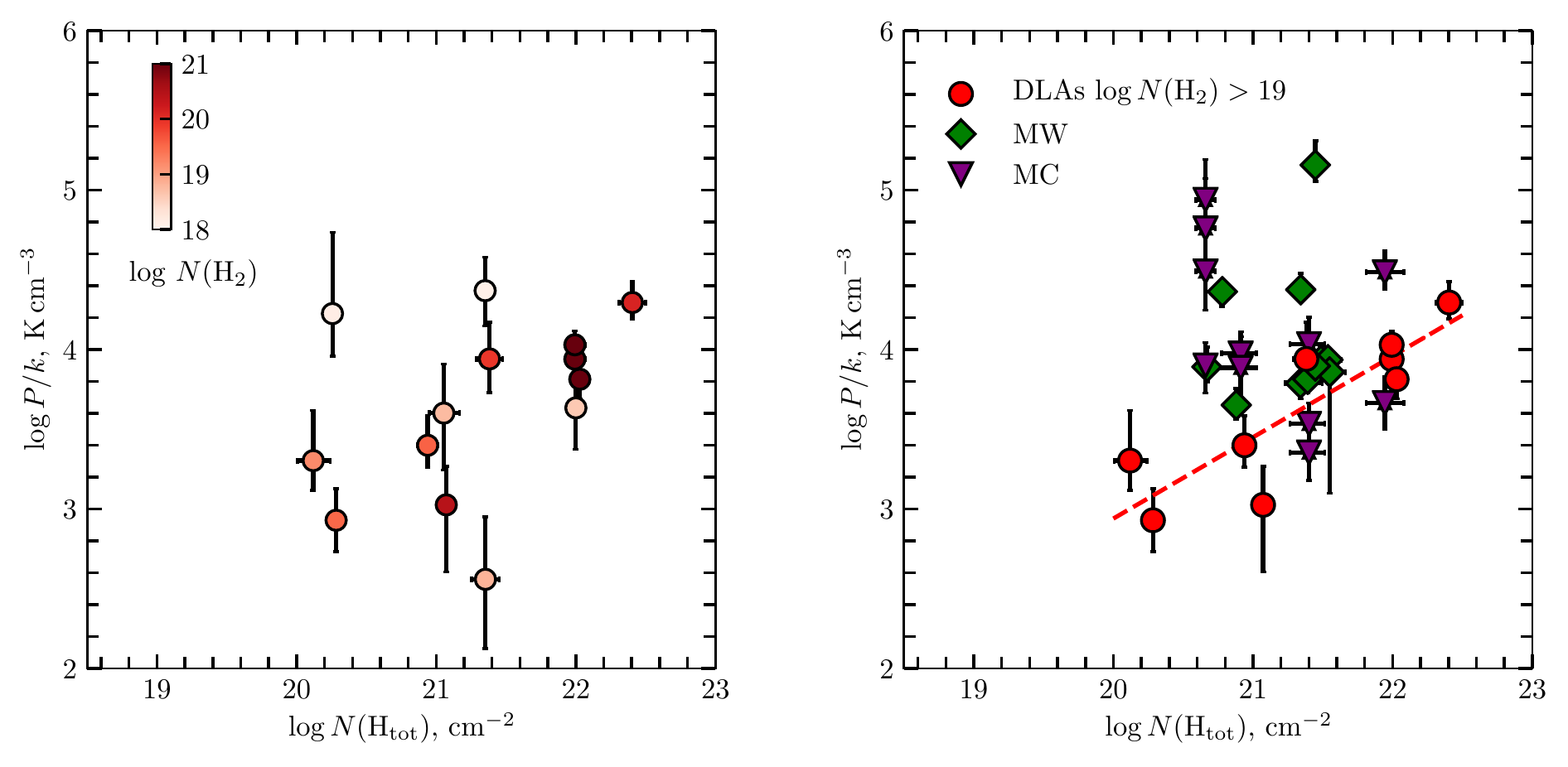}
        \caption{\rm The dependence of the thermal pressure $P/k = n_{\rm H}T$  on total hydrogen column density $\log N({\rm H_{tot}})$. Left panel presents the measurements in high-z H$_2$-bearing DLAs. Color gradient encodes H$_2$ column density. The red, green and purple symbols in the right panel correspond to the measurements in saturated H$_2$ systems with ($\log N>19$) for DLAs and local galaxies (Milky Way and Magellanic Clouds), respectively.}
        \label{pressure_fig}
\end{center}
\end{figure*}

In Fig.\,\ref{pressure_fig} we compare the thermal pressures ($P/k=n_{\rm H}T_{\rm{kin}}$) measured in our samples. 
The pressure in the the $S^{\rm{DLA}}$ sample is in the range $10^{2.5} -10^{4.4}\,\mbox{K\,cm}^{-3}$ with the average about $\log P/k\sim3.6$. 
Similar range of the pressure in study of high-redshift DLAs was found by \cite{Neeleman2015} using excitation of \CII\ and \SiII\ levels ($\log P/k=1.5-5.5$ with mean $3.4$ for 17 DLAs), by \cite{Jorgenson2010} using \CI\ fine structure populations and the \CI/\CII\ ionization balance ($\log P/k=3.9\pm3.7$ for 11 components in 6 DLAs) and recently by \cite{Balashev2019} appling the similar method for sample of 7 H$_2$-bearing DLAs, observed with the VLT/Xshooter ($\log P/k=4.0\pm0.5$). The pressure in the $S^{\rm{MW}}$ and the $S^{\rm{MC}}$ samples are  in the ranges $10^{3.8}-10^{5.2}\,\mbox{K\,cm}^{-3}$ and $10^{3.3}-10^{4.9}\,\mbox{K\,cm}^{-3}$, and average values are respectively $\log P/k\sim4.1\pm0.4$ and $\log P/k =4.1\pm0.5$. Our estimates are higher than ones found in the survey of \CI\ systems in the Milky-Way ($\log P/k =3.58\pm0.18$) by \cite{Jenkins2011} and agree with the estimate of the pressure in the LMC and SMC ($\log P/k =3.56-5.11$) by \cite{Welty2016}. 

Similar to findings in the Section~\ref{sec:T01}, we obtained that for a subsample, $S^{\rm{DLA}}_{19}$, of H$_2$ absorption systems with $\log N({\rm H_2})>19$ there is a strong correlation between the thermal pressure and the total hydrogen column density (see the right panel in Fig.\ref{pressure_fig}). Such correlation was already reported before using one-zone calculations of \CI\ fine-structure excitation \citep[e.g.][]{Balashev2019}. 
The subsample $S^{\rm{DLA}}_{19}$ suggests the following the relation: $\log p_{\th}\simeq0.5\times(\log N({\rm{H}})_{\rm tot}-14.5)$ for $\log N({\rm{H}})_{\rm tot}>20$. We note that $S_{\rm MW}$ and $S_{\rm MC}$ samples do not indicate similar trend.

While we note that the value of H$_2$ column density, $\log N(\rm H_2) > 19$, where aforementioned trends ($I_{\rm UV}/n_{\rm H}$ on $T_{01}$ and $P/k$ on $\log N(\rm H_{\rm tot})$) appear as our primary empirical findings, it is most likely a natural consequence of the H\,{\sc i}-to-H$_2$ transition in the medium. Using PDR models we checked that for the systems from $S^{\rm DLA}$ sample with $\log N(\rm H_2)<19$, the gas in the cloud centers is predominantly atomic\footnote{The position of the the H\,{\sc i}-to-H$_2$ transition depends also on the metallicity and the ratio of $I_{\rm UV}/n_{\rm H}$ \citep{Sternberg2014}. However, for H$_2$-bearing DLA systems from $S^{\rm DLA}$ sample with $\log N(\rm H_2)<19$ we measured similar values of the metallicity of $\sim-1$ and $I_{\rm UV}/n_{\rm H}\sim-1.7$, that gives the transition at $\log N(\rm H_2)\sim19$.} with $f_{\rm H_2} \equiv 2 n({\rm H_2})/n_{\rm H}<0.1$ , while the systems with $\log N(\rm H_2) > 19$ indicate high molecular fractions with $f_{\rm H_2}\gtrsim0.3$ (see an example of the model of J\,0812$+$3208 absorber in the top right panel in Fig.\,\ref{fig01}). That is for absorption systems $\log N(\rm H_2)>19$ the H\,{\sc i}-to-H$_2$ transition is complete, and the thermal state of cold medium probed by these systems may be stabilized by additional cooling by H$_2$ lines. Hence they provide more representative constraints on the thermal pressure. The latter is increased towards the sight lines with higher hydrogen total column densities, as they probe the more central parts of the intervening galaxies \citep{Balashev2017}. This also explains the absence of $P/k-N(\rm H_{tot})$ trend in local galaxies, since both MW and MC samples are not cross-section selected samples, since they used bright stars within the galaxies as background sources. Additionally, MW sample come from the measurement mostly within a galactic plane in the vicinity of the Solar system and the both LMC and SMC samples actually consist only from the couple of sightlines.


Additionally, based on the deeper look into PDR models, we found that the value of $\log N(\rm H_2)>19$ insures us that ortho-para conversion of H$_2$ is complete and the excitation temperature $T_{\rm 01}$ matches the kinetic temperature, that is a direct multiplier in the thermal pressure.

\subsection{Intensity of UV radiation}
\label{sec:uv_intensity}
\begin{figure*}
\begin{center}
        \includegraphics[width=1.0\textwidth]{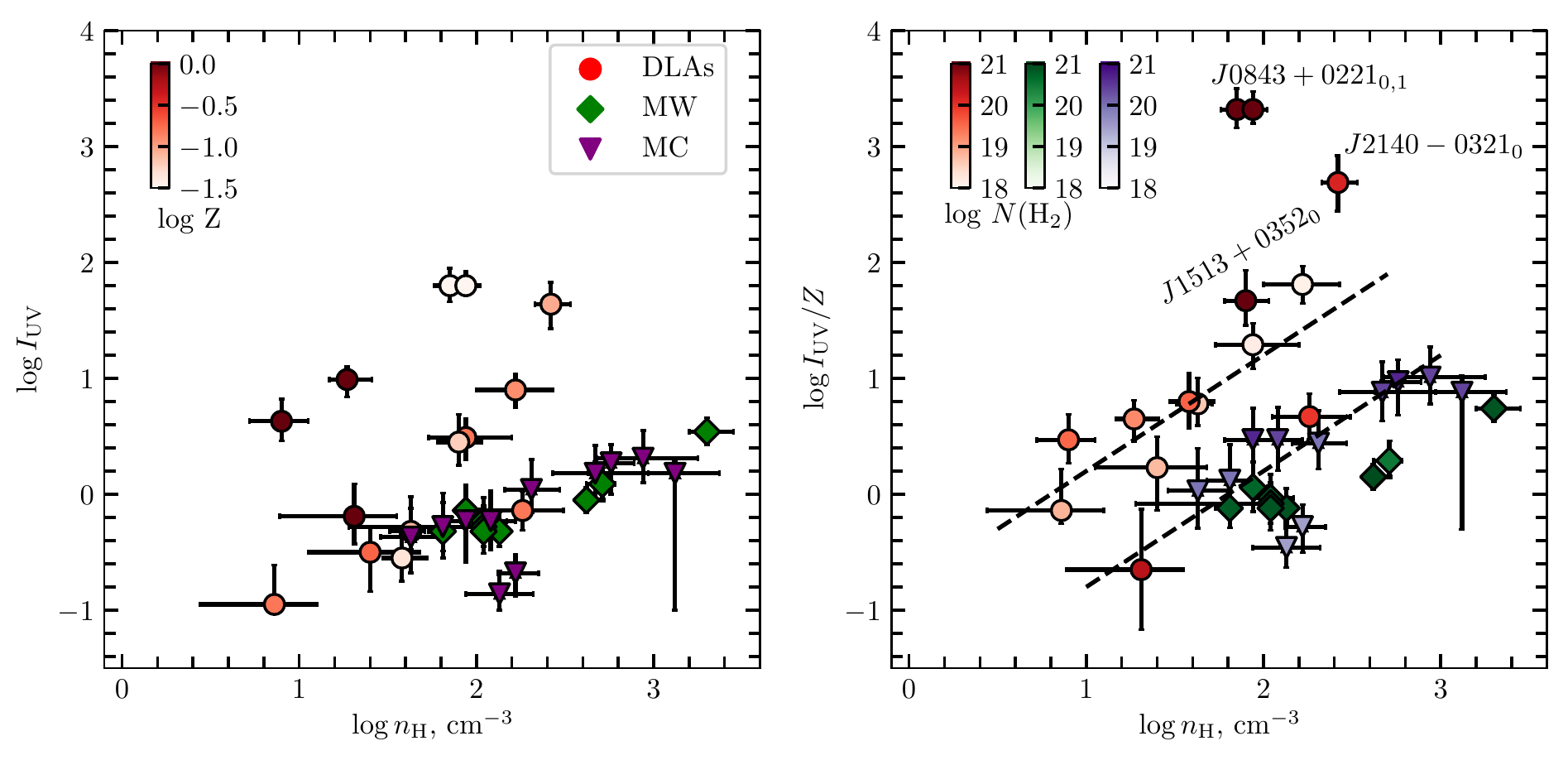}
        \caption{\rm 
        {Left and right panels respectively show the dependence of the UV intensity and the ratio of UV intensity to metallicity on the number density. The red circles, green diamonds and purple triangles represent values derived in high redshift DLAs and H$_2$ systems in the Milky-Way and Magellanic Clouds, respectively. The color gradients in the left and right panels encode the metallicity and H$_2$ column density, respectively. The dashed lines on the right panel represent a linear dependence of the $I_{\rm UV}/Z$ on the hydrogen number density. Color gradient encodes the metallicity of the DLA sample in the left panel and the column density of H$_2$ for the samples of DLAs, Milky-Way and Magellanic Clouds in the the right panel.} 
        }
        \label{metallicity_fig}
\end{center}
\end{figure*}

The intensity of the incident UV radiation and number density can be well measured by our method owing to the orthogonal dependence of regions in $I_{\rm UV}-n_{\rm H}$ plane, constrained by H$_2$ and \CI\ excitation. 
The estimates of UV intensity for the $S^{\rm DLA}$ and local ($S^{\rm MW}$ and $S^{\rm MC}$) samples are given in Table\,\ref{list} and Table \ref{listMW}, respectively. 
We found that the UV intensity in the $S^{\rm DLA}$ sample is varied in the wide range of $0.1-100$ units of the Mathis field. On the contrary, the systems in the  $S^{\rm{MW}}$ and  $S^{\rm{MC}}$ samples are characterized by low values ranges $\sim0.1-3$ units of Mathis field. The results are shown in Fig.\,\ref{metallicity_fig}.

There is a clear trend that in systems with higher UV intensity we detect higher hydrogen density $n_{\rm H}$, which naturally consequence of the thermal balance (see also Sect.~\ref{sec:T01}). Indeed studied systems in the samples  have temperatures $\sim100$\,K, which evidently lead to the linear dependence of $I_{\rm UV}$ on $n_{\rm H}$ (see Eq.\,\ref{eq_th_balance}). However, $S^{\rm DLA}$ sample indicates a significant dispersion around this linear trend, some systems shown systematically higher $I_{\rm UV}$ (see left panel in Fig.\,\ref{metallicity_fig}). There may be different explanation of the high observed dispersion of UV intensities. However, the most straightforward one is that the $S^{\rm DLA}$ sample corresponds to the galaxies at redshift $z \sim 2-3$, where interstellar UV field is expected to be higher, since star-formation rate is peaked at these redshifts (see, e.g. \citealt{Madau2014}). However, one should bear in mind that as a whole DLA systems (due to their selection by cross-section) probed mostly low-mass galaxies (i.e. probably not starbursts ones), that are dominated by number in galaxy population at particular redshift. Additionally the $S^{\rm DLA}$ sample is not typical DLAs, since only a small fraction of DLAs ($4\pm1$\%, see \citealt[][]{Balashev2018}) shows H$_2$ and hence \CI\ absorptions \citep{Noterdaeme2018} 
Therefore our measurement can indicate that UV field is statistically enhanced in the cold phase of ISM of high-$z$ galaxies. 

On the other hand one can note that $S^{\rm DLA}$ sample probes wide ranges of the metallicities. Actually, observed high dispersion in $S^{\rm DLA}$ around expected linear dependence of $I_{\rm UV}$ on $n_{\rm H}$ can be reduced by correction for the metallicity. Indeed, since H$_2$ is formed on the dust grains, H$_2$ formation rate is scaled with metallicity, while H$_2$ destruction rate is linearly proportional to $I_{\rm UV}$ \citep[see e.g.]{Bialy2016}. It means that at low metallicities we should have lower UV flux to form H$_2$ at some characteristic number density. After correction for metallicity (e.g. considering $I_{\rm UV}/Z$ versus $n_{\rm H}$ plane, see the right panel of Fig.\,\ref{metallicity_fig}), the dispersion in the $S^{\rm DLA}$ sample around linear dependence of $I_{\rm UV}/Z$ on $n_{\rm H}$ is significantly reduced, except two systems ($J0843+0221$ and $J2140+0321$), which are extremely saturated DLAs (ESDLAs, with $\log N({\rm{HI}})\ge21.7$, \citealt{Noterdaeme2014}). These ESDLA systems are reasonably biased towards higher UV fluxes, since they most likely probe central parts of the remote galaxies (i.e. at very small impact parameters, see \citealt{Ranjan2018, Ranjan2020}), where UV flux, can be significantly enhanced on average, and locally due to higher density of star-forming regions.    
One can expect that such systems should be more similar to ISM clouds observed in disk of Milky-Way or Magellanic Clouds. However we see that they exhibit a higher UV field intensity, thus  emphasizing the difference  between local and high-$z$ galaxies. Whether this is a common feature or a statistical outlier, it may become clear once more observations observations of such systems will be available.


\section{Conclusion }
\label{sec:conclusion}

We present a systematic study of physical conditions in the cold H$_2$-bearing ISM of high redshift DLAs and in the Milky-Way and Magellanic Clouds galaxies. Our measurements based on the fit to the observed populations of H$_2$ rotational levels and fine-structure levels of \CI\ in known absorption systems, detected in spectra of high redshift quasars and stars in local galaxies. 

The modelling of H$_2$ levels population requires an accurate calculation of the radiative transfer, and therefore only in several specific cases the detailed analysis of H$_2$ and \CI\ excitation has been done. To analyse all known H$_2$/\CI\ systems we calculated grids of constant-density model, which were uniformly distributed in the space of three main physical parameters -- the metallicity, hydrogen density and intensity of UV field. For modelling H$_2$ and \CI\ level populations we used the PDR Meudon code \citep{LePetit2006}, which performs a complete calculation of the radiative transfer of UV radiation in the UV lines of H$_2$ in combination with a solution of the thermal balance and chemistry.
We found that in many cases a joint analysis of low H$_2$ rotational levels and \CI\ fine-structure levels allows one to break the degeneracy in the $I_{\rm UV}-n_{\rm{H}}$ plane and provides significantly tighter constraints on the number density and intensity of UV field. 

We applied this method to analyse physical conditions in the samples of twelve strong H$_2$-bearing high redshift DLAs and fourteen \CI-bearing H$_2$ absorption systems in the Milky-Way and Magellanic Clouds. We found that H$_2$-bearing gas in these systems is cold with typical kinetic temperature $\sim100$\,K and dense with the number density $10-500$\,cm$^{-3}$. 
The values of the temperature, number density and thermal pressure are in good agreement between these three samples. However, we found that the intensity of UV field in the sample of DLAs is varied in wide range $0.1-100$ units of Mathis filed, while it is $\sim0.1-3$ units in the samples of H$_2$ systems in the Milky-Way and Magellanic Clouds. The large dispersion and measured values of UV field in DLAs sample is naturally expected, since it probes the population of the distant galaxies at redshifts $z>2$, where the interstellar UV field can be higher, due to the peak of star-formation at these redshifts. We found that for $S^{\rm DLA}$ sample the dispersion around dependence of $I_{\rm UV}/Z$ on $n_{\rm H}$ is significantly less than dispersion around dependence of $I_{\rm UV}$ on $n_{\rm H}$. This most likely linked with the scaling of H$_2$ formation rate with these parameters.

For a subsample of H$_2$ absorption systems with $\log N(\rm H_2)>19$ we found a linear trend of increase of UV field intensity with increasing number density and confirm the trend of increase of thermal pressure with total hydrogen column density. The first trend is naturally expected from the ISM thermal balance, while the second trend confirm the earlier findings that the higher column density DLAs probe the central parts of the remote galaxies, where the thermal pressure is enhanced.
Our findings indicate that the study of the H$_2$/\CI-bearing DLAs is very promising tool to get insights of the physical state of the cold diffuse ISM of local and high-redshift galaxies.

\section{Data availability}
The paper is based on the published measurements of the H$_2$/\CI-bearing absorption systems at high redshifts and local galaxies.

\section*{Acknowledgements}

This work is partially supported by RFBR 18-32-00701. SB is partially supported by Basis foundation.

\appendix

\section{Detailed fit}
\label{ApppxA}
In this section we present fit to observed populations of rotational levels of H$_2$ and fine-structure levels of \CI\ in the $S^{\rm{DLA}}$, $S^{\rm{MW}}$ and $S^{\rm{MC}}$ samples. In Figs\,\ref{supples0}-\ref{Mag_part2} we show constraints on the hydrogen number density $n_{\rm{H}}$ and UV field strange (in left panels) and best fit to population of H$_2$ rotational levels (middle panel) and relative populations of excited \CI\ fine-structure levels (right panels). Systems are  arranged in increasing order of coordinates. 
 
\label{suppls}

\begin{figure*}
\begin{center}
        \includegraphics[width=\textwidth]{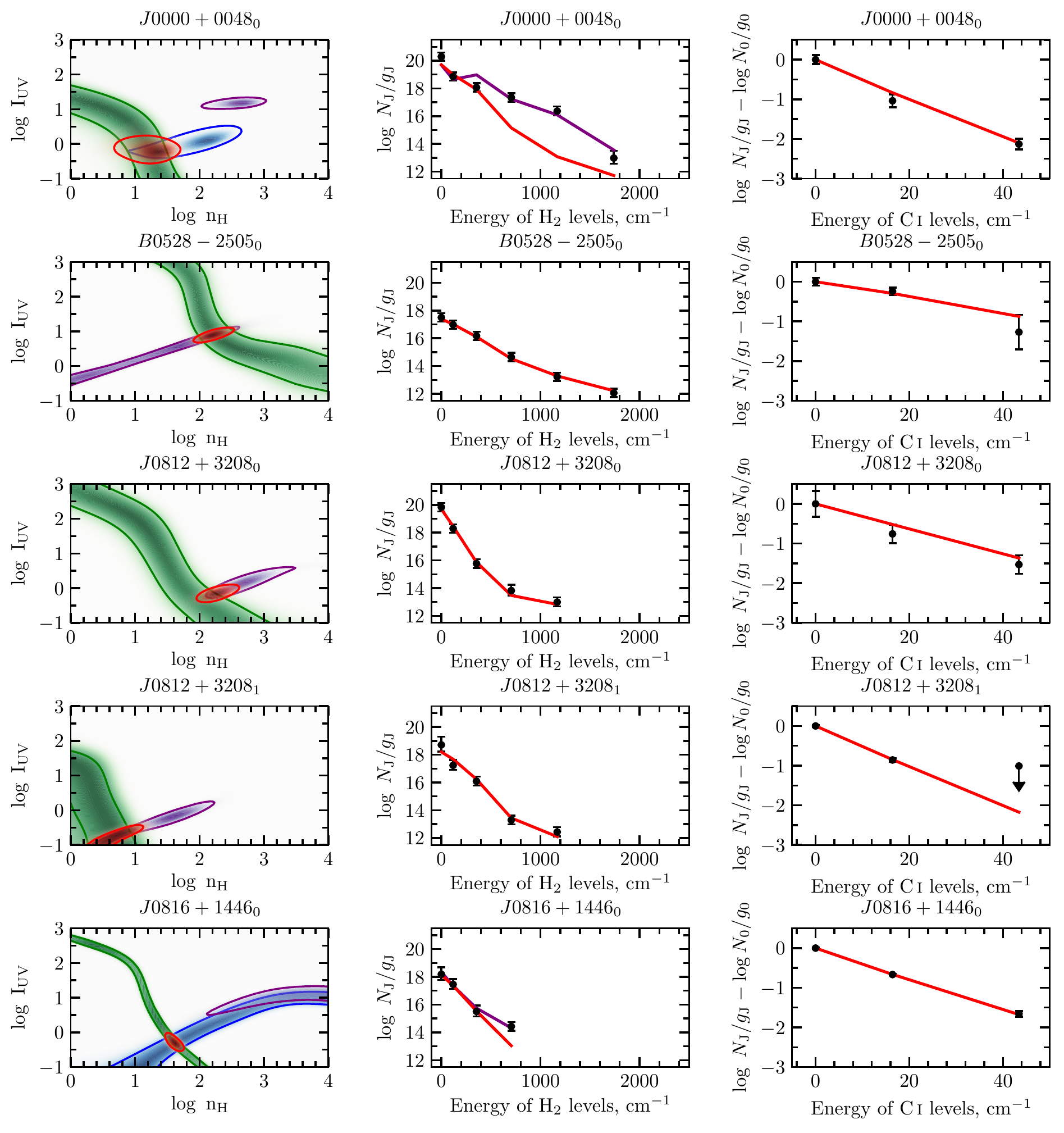}
        \caption{\rm Fit to observed populations of H$_2$ and \CI levels in the $S^{\rm DLA}$ sample. In the left panel the green and purple contours represent constraints on the hydrogen number density and UV radiation intensity, obtained with the analysis of excitation of \CI\ fine-structure and  H$_2$ rotational levels (all observed), respectively. In some cases, we also show constraint obtained with the fit to excitation of lower rotational levels of H$_2$ (J=0 to J=2) by the blue contour. Red contour represents a joint constraint using \CI\ and (lower) H$_2$ levels (if the blue contour is presented). The middle and right panels show the population of H$_2$ rotational and \CI\ fine-structure levels, respectively. The black circles indicate the observed values. The red lines correspond to the join fit to population of both H$_2$ and \CI\ levels. The purple lines, presented in some middle panels, show the best fit obtained using only all observed H$_2$ rotational levels.}
        \label{supples0}
\end{center}
\end{figure*}

\begin{figure*}
\begin{center}
        \includegraphics[width=\textwidth]{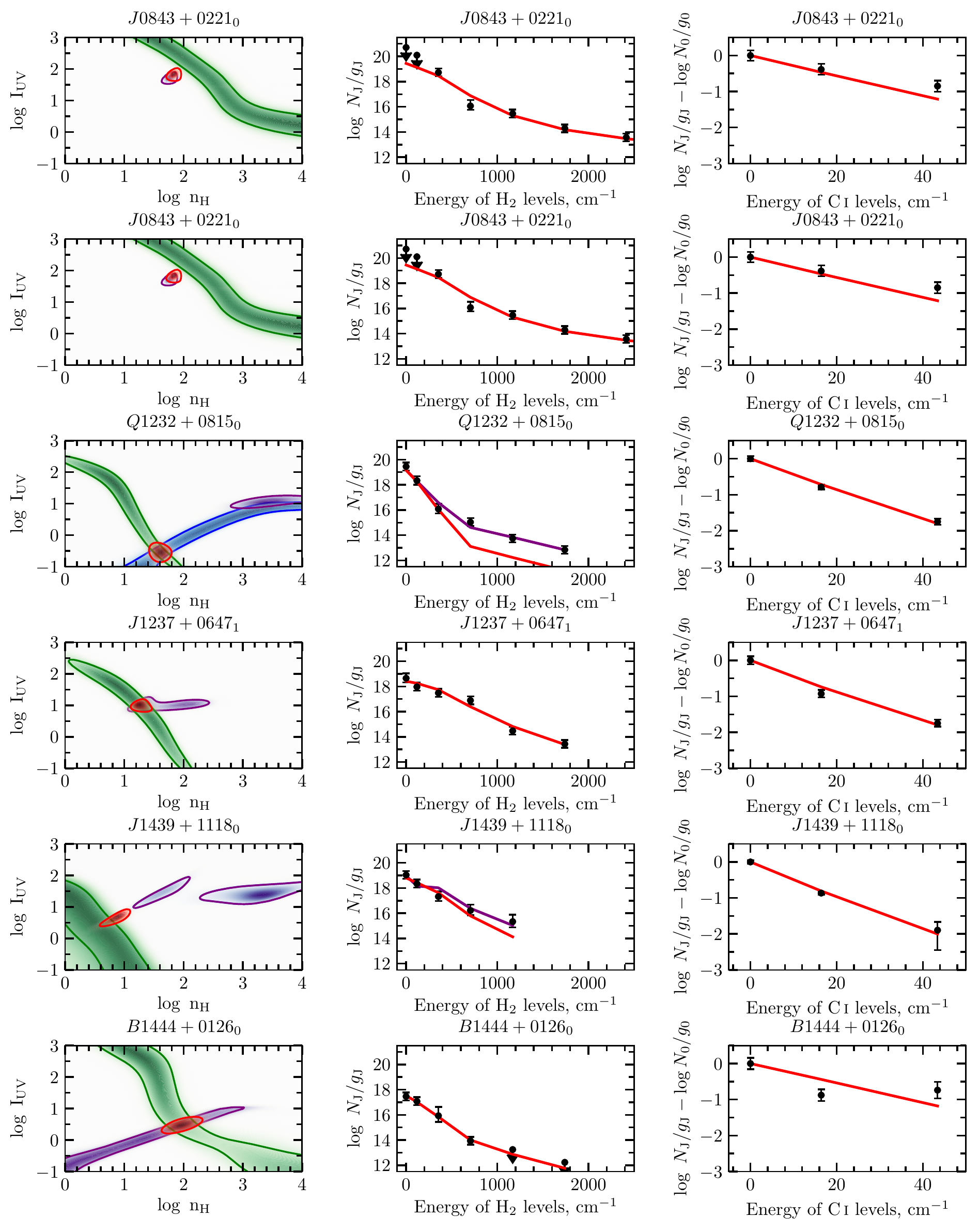}
        \caption{Continuation of Fig.~\ref{supples0} \label{supples1}} 
\end{center}
\end{figure*}

\begin{figure*}
\begin{center}
        \includegraphics[width=\textwidth]{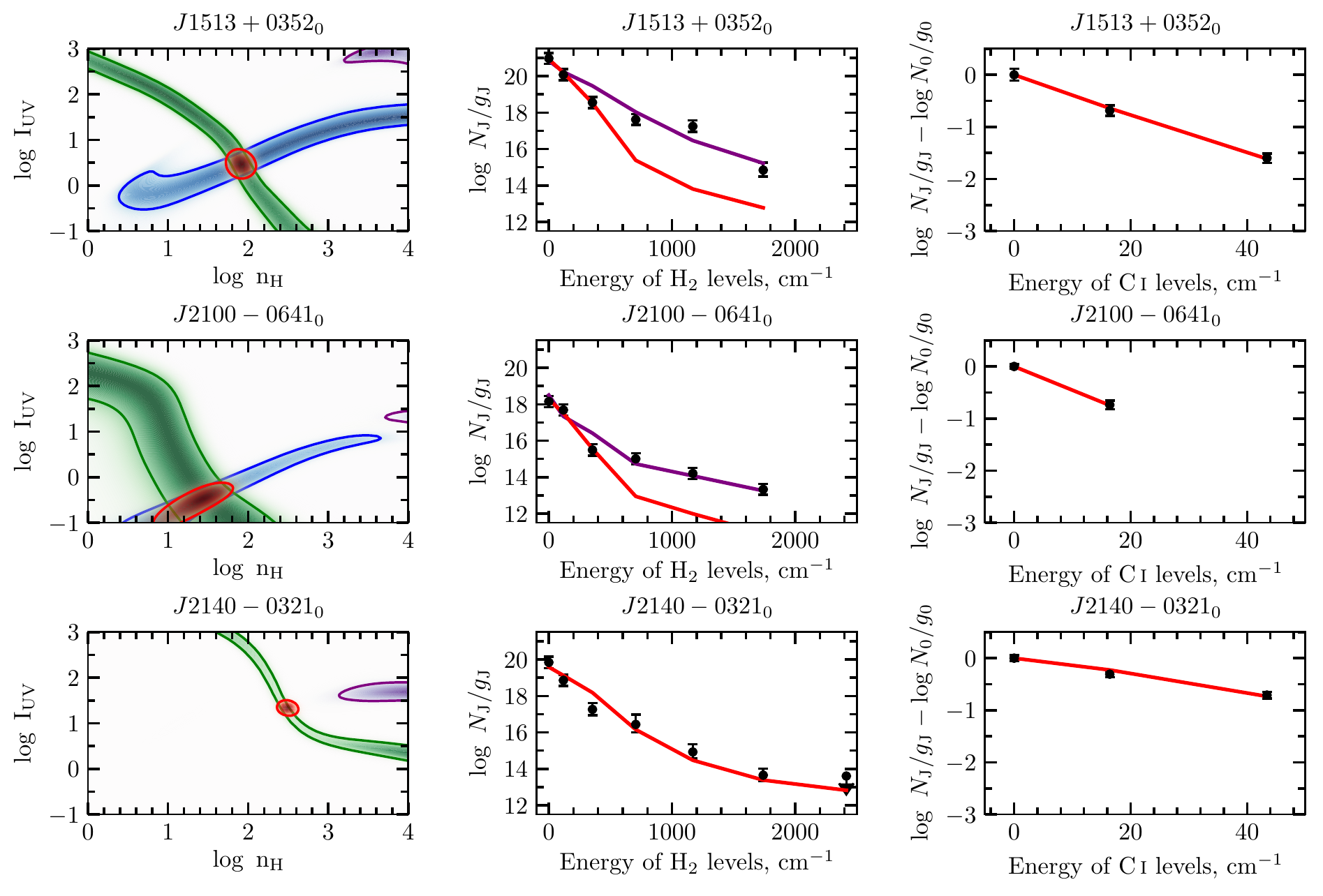}
        \caption{Continuation of Fig.~\ref{supples0}
        \label{supples2}
        }
\end{center}
\end{figure*}


\begin{figure*}
\begin{center}
        \includegraphics[width=\textwidth]{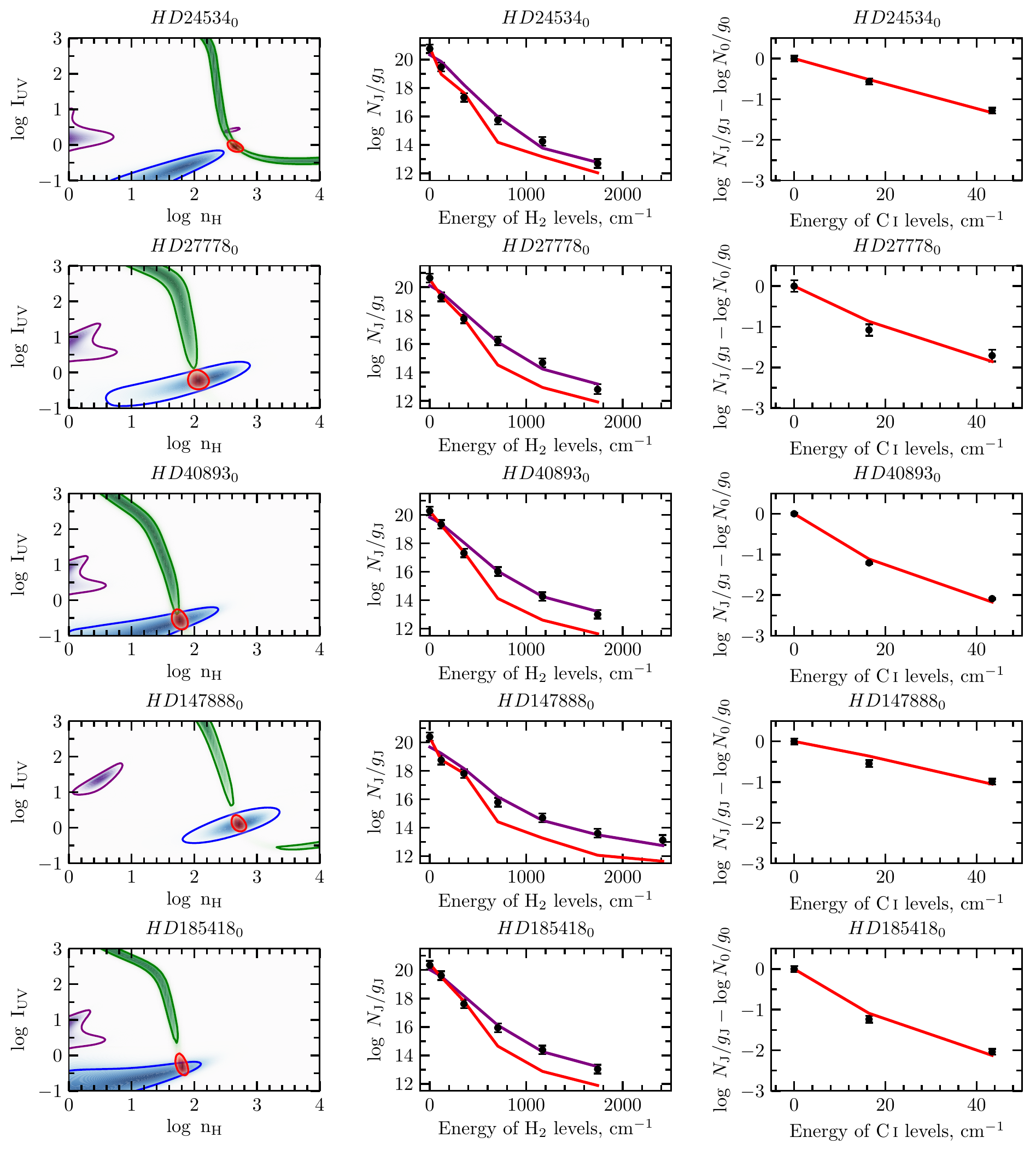}
        \caption{\rm Fit to populations of H$_2$ and \CI levels in the sample of H$_2$ absorption systems observed in the Milky-Way. The contours, lines and color encoding are the same as in Fig.~\ref{supples0}.}
        \label{MW_part1}
\end{center}
\end{figure*}

\begin{figure*}
\begin{center}
        \includegraphics[width=\textwidth]{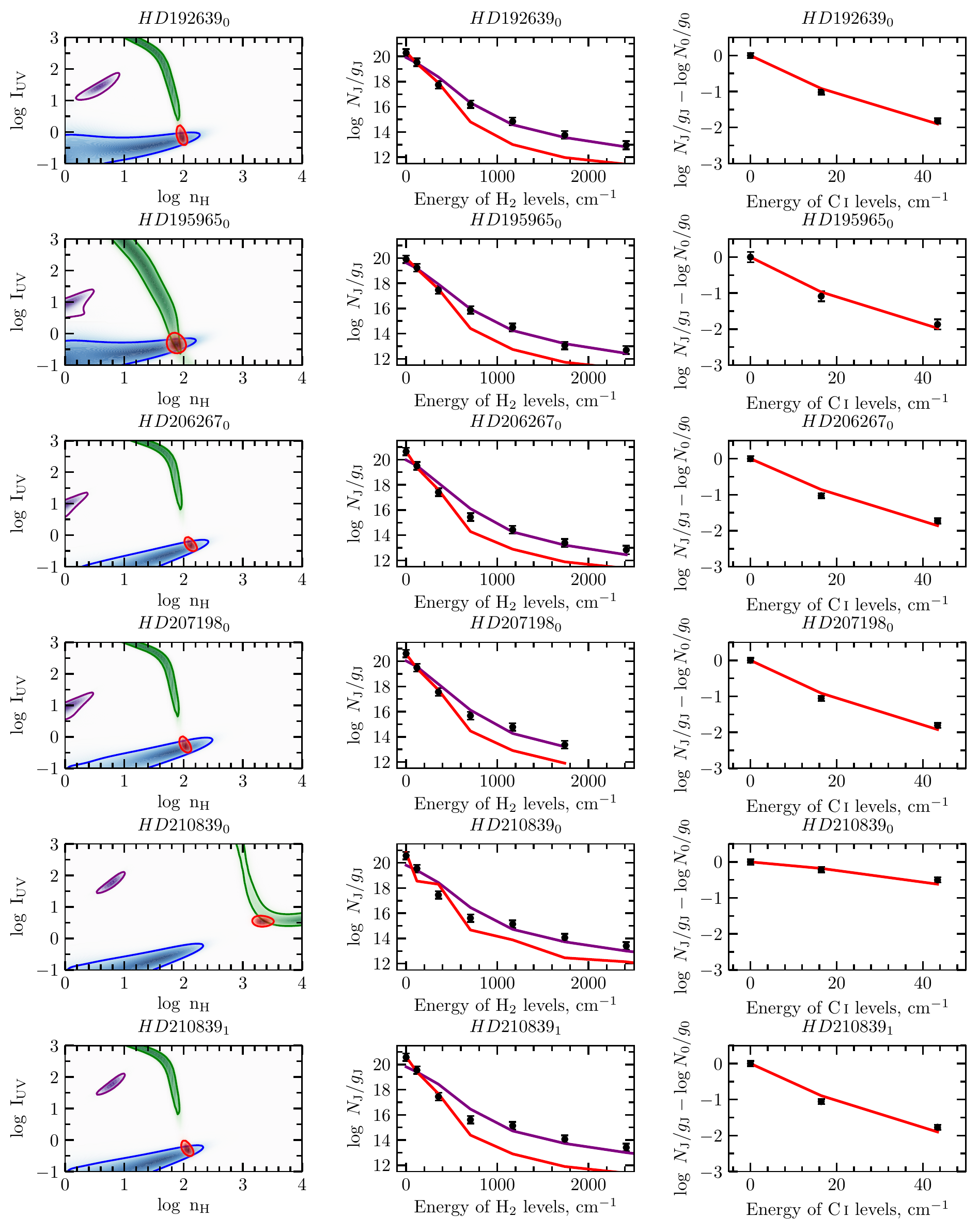}
        \caption{\rm Continuation of Fig.~\ref{MW_part1}}
        \label{MW_part2}
\end{center}
\end{figure*}

\begin{figure*}
\begin{center}
        \includegraphics[width=\textwidth]{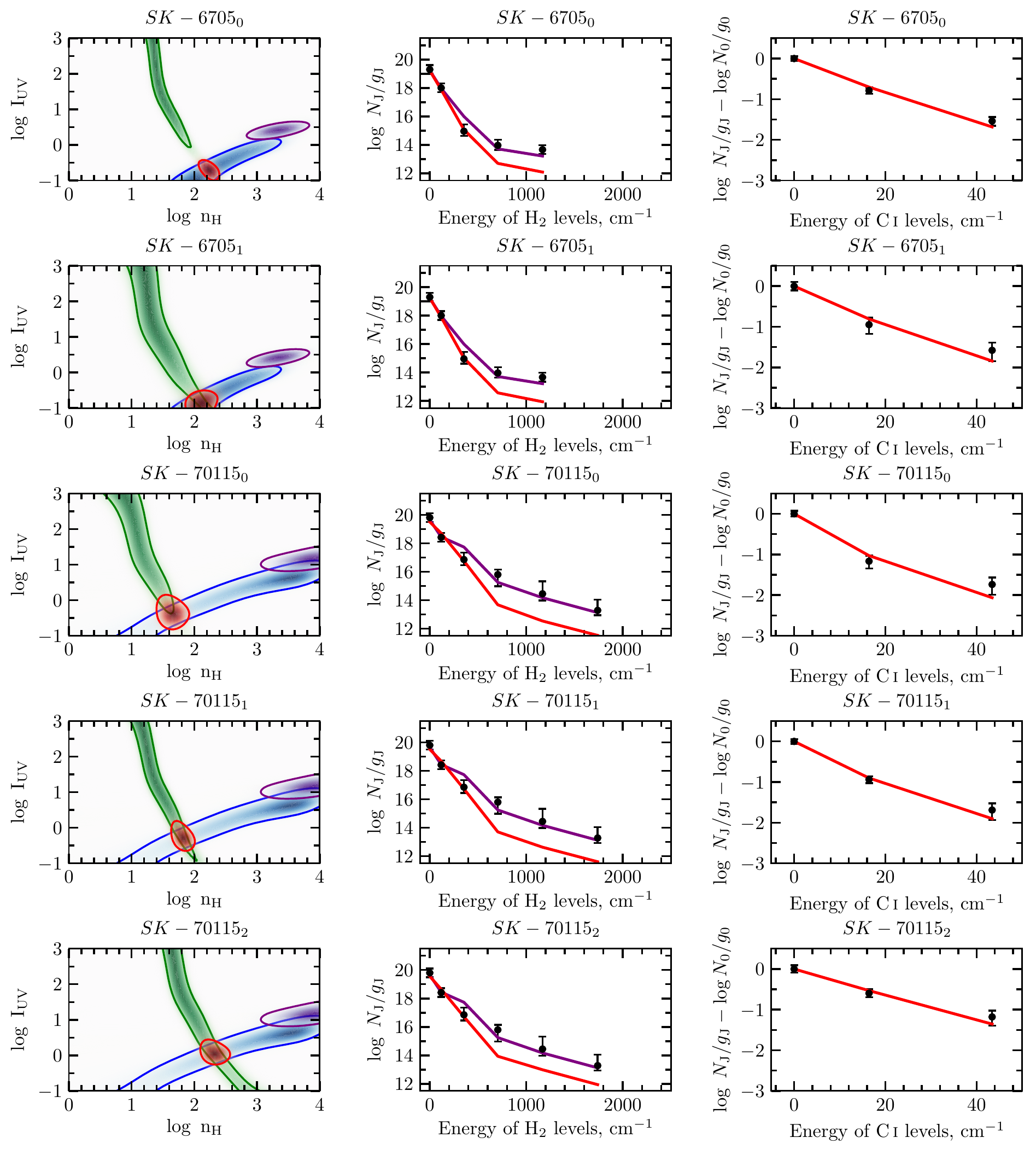}
        \caption{\rm Fit to populations of H$_2$ and \CI levels in the sample of H$_2$ absorption systems observed in the Large Magellanic Cloud. The contours, lines and color encoding are the same as in Fig.~\ref{supples0}.}
        \label{Mag_part1}
\end{center}
\end{figure*}

\begin{figure*}
\begin{center}
        \includegraphics[width=\textwidth]{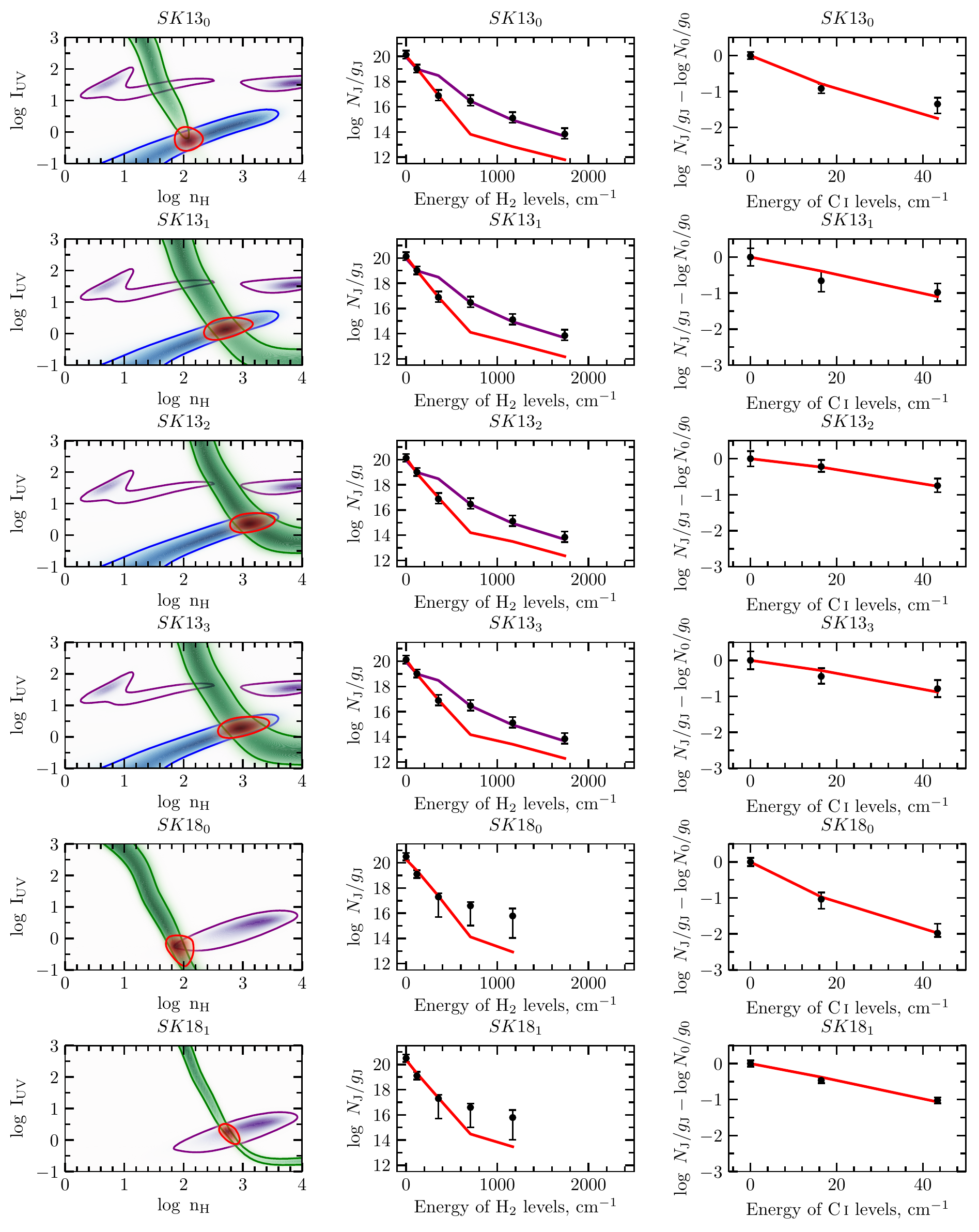}
        \caption{\rm Fit to populations of H$_2$ and \CI levels in the sample of H$_2$ absorption systems observed in the Small Magellanic Cloud. The contours, lines and color encoding are the same as in Fig.~\ref{supples0}.}
        \label{Mag_part2}
\end{center}
\end{figure*}



\bibliographystyle{mnras}
\bibliography{library} 
\bsp	
\label{lastpage}
\end{document}